\documentclass[pre,amsmath,amssymb,floatfix]{revtex4}
 
\usepackage{graphicx}
\usepackage{dcolumn}
\usepackage{bm}
\usepackage{epsf}

\usepackage{graphics}

\usepackage{amssymb}

\begin{document}


\title{The Integral Equation Method for a Steady Kinematic Dynamo Problem}
\author{M. Xu, F. Stefani, and G. Gerbeth}
\affiliation{Forschungszentrum Rossendorf, P.O. Box 510119, 
D-01314 Dresden, Germany}

\begin{abstract}

With only a few exceptions, the numerical simulation of 
cosmic and laboratory hydromagnetic dynamos 
has been 
carried out in
the framework of the differential equation method.
However, the integral 
equation method is  known to provide robust and accurate 
tools for the numerical solution of many problems in other fields 
of physics.
The paper is intended to facilitate the use of integral equation 
solvers in dynamo theory.
In concrete, the integral equation method is employed to solve 
the eigenvalue
problem for a hydromagnetic dynamo model with an 
isotropic helical turbulence parameter $\alpha$. 
For the case of spherical geometry, 
three examples of the function 
$\alpha(r)$ with steady and oscillatory solutions are 
considered. A convergence rate 
proportional to the inverse squared of the number of grid points 
is achieved. Based on this method, a convergence accelerating strategy 
is developed and the convergence rate is improved remarkably. 
Typically, quite accurate results can be obtained
with a few tens of grid points.
In order to demonstrate its suitability for the treatment of dynamos in other than
spherical domains, the method is also applied to 
$\alpha^2$ dynamos in rectangular boxes. The
magnetic fields and the electric potentials for the first eigenvalues
are visualized. 
\end{abstract}

\maketitle

\section{Introduction}

The hydromagnetic dynamo effect is the cause of the magnetic fields
of planets, stars, and galaxies \cite{KRRA,MOFF}. In the last 
decades much progress
has been made in the analytical and numerical understanding
of magnetic field generation in cosmic bodies. Only 
recently, the homogeneous dynamo effect has been validated 
experimentally
in large liquid sodium facilities in Riga and Karlsruhe 
\cite{GAI4,MUST,GAI5,STMU,GAI6}.

The usual numerical method to simulate hydromagnetic dynamos
is based on the differential equation method. In the case of 
{\it kinematic
dynamo models}, for which the fluid velocity $\bf{v}$ is supposed to be
given and unchanged, the relevant differential equation is the 
induction equation for the magnetic field $\bf{B}$,
\begin{eqnarray}
\frac{\partial {{\bf{B}}}}{\partial t}={\bf \nabla}
\times ({\bf{v}} \times {\bf{B}})
+\frac{1}{\mu_0 \sigma} \Delta {\bf{B}} \label{4} \; ,
\label{eq0}
\end{eqnarray}
with $\mu_0$ and $\sigma$ denoting the permeability of the free space 
and the electrical 
conductivity 
of the fluid, respectively. Note that the
magnetic field has to be source-free:
\begin{eqnarray}
{\bf \nabla} \cdot {\bf{B}}=0 \label{0a} \; .
\end{eqnarray}
Assuming that there are no external excitations of the            
magnetic field from outside the considered finite region,          
the boundary condition for the magnetic field reads                
\begin{eqnarray}                                                  
{\bf{B}}={\bf{O}}(r^{-3}) \; \;\; \mbox{as}                         
\; \;\; r \rightarrow \infty \label{3a} \; .                       
\end{eqnarray}

For a qualitative understanding of Eq. (\ref{eq0}) one 
should notice
that the magnetic field evolution 
is governed by the competition between the
diffusion and the advection
of the field. Without the advection term
the magnetic field would disappear within a typical decay
time $t_d= {\mu_0 \sigma} l^2$,
with $l$ being a typical length scale of the system.
The advection can lead to an increase of
${\bf{B}}$ within a kinematic time $t_k=l/v$. If the
kinematic time
becomes smaller than the diffusion time, the net effect of
the evolution can become positive, so that magnetic field self-excitation
can start.     
Relating the diffusion time-scale to the kinematic time-scale
we get a dimensionless number that governs the evolution of the
magnetic
field. This number is called the magnetic Reynolds number $R_m$:
\begin{eqnarray}
R_m=\mu_0 \sigma l v \; .
\end{eqnarray}
Depending on the particular flow pattern,                                  
the values of the critical $R_m$, at which self-excitation occurs, 
are in the                        
range of 10$^1$...10$^3$.                                              

For the more complicated  case 
of {\it dynamically 
consistent dynamo models} one has to solve simultaneously the induction equation
for the magnetic field and the Navier-Stokes 
equation for the velocity, in which the back-reaction of the Lorentz forces 
on the flow
has to be included.

A considerable part of dynamo research has been devoted 
to magnetic field
self-excitation in finite spherical bodies, such as the Sun or the Earth.
Fortunately, for the spherical case the boundary conditions for 
the magnetic 
field can be formulated separately for every degree and order of the 
spherical harmonics, so that the treatment of the magnetic fields in 
the exterior 
can be avoided.

This pleasant situation changes drastically when dynamos in other
than spherical domains are considered. Then the correct 
treatment of
the boundary conditions becomes non-trivial. 
In particular, this problem arises in connection
with galactic magnetic fields, and 
in simulations related to the
recent dynamo 
experiments \cite{GAI4,MUST,GAI5,STMU,GAI6} which are carried
out in cylindrical vessels. 
There are three ways to 
circumvent this problem:
\begin{itemize}
\item The correct non-local boundary conditions are replaced by simplified
local boundary conditions, e.g. ``vertical field conditions'' or 
``pseudo-vacuum boundary conditions'' \cite{BRAN,RUZH}, 
demanding that the magnetic field has only a normal 
component at the boundary. This method is  very cheap from the 
numerical point of view, but it is of course not correct.
\item The real dynamo body is virtually embedded into a larger 
sphere for which the
well-known boundary conditions for every degree and order of the 
spherical harmonics 
can be
used. The region between the real dynamo  and the 
surface of the
virtual sphere is thought to be filled by a medium with a lower 
conductivity 
than that of the dynamo domain. Scaling this 
artificial
conductivity to lower and lower values one can look for 
the convergence of the
results. This method was successfully employed for the simulation of the
Karlsruhe dynamo experiment \cite{RAE1,RAE2}, where 
the dynamo module has an aspect ratio (ratio of height to radius)
close to one.
\item The Laplace equation for the magnetic field is 
solved in the exterior of the dynamo domain, and the 
interior solution is
matched to the exterior solution by  using the correct 
boundary conditions.
This method, which was used for the simulation of the 
Riga dynamo experiment \cite{STE1},
is correct but numerically expensive.
\end{itemize}

This unsatisfactory situation concerning the handling
of boundary conditions
was our main motivation to
reconsider the integral equation method to dynamos in finite
domains \cite{STE2}. The 
formulation of this method for the case of steady dynamos, 
which is nothing other than the application of 
Biot-Savart's law to dynamos, 
can already be found in the book of Roberts \cite{ROB1}. 
Interestingly enough, in Roberts opinion (\cite{ROB1}, p. 74)
this formulation did 
``...not appear, in general, to be very useful.'' 
The integral equation 
method was 
used in a few previous papers \cite{GAI2,GAI3,FREI,DORA}, 
in which the effect of boundaries 
was mostly discarded, however.
The ''velocity-current-formulation'' 
by Meir and Schmidt \cite{MES2} was intended 
to circumvent the numerical treatment outside the region of
interest. However, the numerical focus of this work laid more on 
coupled MHD problems with  small magnetic Reynolds number
than on dynamo problems. 

A concrete result of our recent paper \cite{STE2}
was the formulation of 
a system of one-dimensional integral equations for a dynamo model with a
spherically symmetric, isotropic helical turbulence parameter $\alpha$
in a finite
sphere, and the 
re-derivation of the solution found by Krause and Steenbeck \cite{KRST} for
the special case of constant $\alpha$.

This system of integral equations 
for the case of spherically symmetric, isotropic
$\alpha$ is also at the root 
of the first numerical examples considered in this paper.
Our present goal is  to study and optimize the 
performance of numerical schemes to solve the integral
equations for dynamos of this sort. 
The restriction to spherically symmetric $\alpha$ has the advantage that 
the equations decouple for every degree and order of the spherical
harmonics. That makes our method comparable to
the corresponding integral equation method for the radial 
Schr\"odinger equation \cite{BUEN,GON1,GON2}.
From there, and from other applications of the integral equation method
\cite{GRRO,ATKI,HACK,DELV1,PTVF}, it is well known that the linear 
systems arising
from the discretization of integral 
equations are generally well-conditioned.
We present the numerical results of an integral equation solver with
a convergence rate proportional to the inverse squared of the number 
of grid points. We also show how the convergence can be improved 
drastically by 
using a convergence accelerating strategy. 

Whereas these examples for the case of spherical geometry illustrate 
the feasibility of the integral equation approach and its equivalence
with the differential equation approach they do not demonstrate any
particular improvement with respect to the latter. The main advantage 
of the integral equation
approach, its suitability for the treatment of dynamos in arbitrary domains,
is therefore exemplified by another example which 
would be very hard to deal within the
differential equation approach. 
Again, we consider an $\alpha^2$ dynamo, but restrict the electrically 
conducting and
dynamo active domain 
to a rectangular box outside which we assume vacuum.
For such ''matchbox dynamos'', we compute the first eigenvalues and visualize the 
magnetic field and the electric potential structure.
It is shown how the first three eigenvalues, which are different for the 
case of different side lengths of the box, converge for the case of a cubic 
box.

\section{Basics}

In this section, we compile the necessary formulae
which are at the root of our numerical investigation.
For details of the derivation we refer to our previous
paper \cite{STE2}.

Basically, our considerations are restricted to the steady case, 
i.e., to dynamos
with growth rate {\it and} frequency equal to zero. 
Quite generally \cite{KRRA}, the electromotive force (emf) in turbulent 
flows of conducting fluids can be written
in the form
\begin{eqnarray}
\mathbf{F}=
  \mathbf{v} \times \mathbf{B}
+\alpha \mathbf{B} -\beta {\bf \nabla} \times \mathbf{B} \; .
\label{eq1}
\end{eqnarray} 
The first term in this equation is the usual emf induced 
in a fluid flowing with the mean velocity $\mathbf{v}$ under the 
influence of the magnetic field $\mathbf{B}$. The second term, 
$\alpha \mathbf{B}$, represents the effect of a helical
turbulence, with $\alpha$ characterizing the 
helical part of the turbulence that can be derived  in the framework 
of mean-field magnetohydrodynamics \cite{KRRA}. 
The concept of the $\alpha$-effect 
plays a considerable role in dynamo theory.
The term  $\beta {\bf \nabla}  \times \mathbf{B}$  reflects
the decrease of the electrical conductivity due to turbulence.

In the steady case, the 
current density 
$\mathbf{j}$ is given by
\begin{eqnarray}
\mathbf{j}=\sigma (\mathbf{F}-{\bf \nabla} \varphi) \; ,
\label{eq2}
\end{eqnarray}
where $\sigma$ is the conductivity of the fluid and
$\varphi$ is the electrostatic potential.
With these notations,
the coupled system of integral equations
for steady kinematic dynamos can be written in the following form
\cite{ROB1,STE2}:
\begin{eqnarray}
{\mathbf{B}}({\mathbf{r}})&=&\frac{\mu_0 \sigma}{4 \pi} \int_D
\frac{ {\mathbf{F}}  ({\mathbf{r'}}) \times 
({\mathbf{r}}-{\mathbf{r'}})}{|{\mathbf{r}}-{\mathbf{r'}}|^3} \; 
dV'-\frac{\mu_0 \sigma}{4 \pi} \int_S \varphi({\mathbf{s'}}) 
{\mathbf{n}} ({\mathbf{s'}}) \times
\frac{{\mathbf{r}}-{\mathbf{s'}}}{|{\mathbf{r}}-{\mathbf{s'}}|^3} 
\; dS'\label{eq3} \\
\varphi({\mathbf{s}})&=& 
\frac{1}{2 \pi} \int\limits_D  
\frac{
{\mathbf{F}}({\mathbf{r'}}) \cdot ({\mathbf{s}}-{\mathbf{r'}})}{|{\mathbf{s}}-
{\mathbf{r'}}|^3} 
\; dV' -\frac{1}{2 \pi} \int\limits_S \varphi({\mathbf{s'}})
{\mathbf{n}}({\mathbf{s'}}) 
\cdot \frac{{\mathbf{s}}-{\mathbf{s'}}}{{|{\mathbf{s}}-{\mathbf{s'}}|}^3} 
\;  dS' \; ,
\label{eq4}
\end{eqnarray}
with $\mu_0$ being the magnetic permeability of the 
free space, $\mathbf{n}({\mathbf{s'}})$ denoting the outward directed 
unit vector 
at the boundary point $\mathbf{s'}$, and $dS'$ denoting an area element
at this point. 
$D$ and $S$ indicate integrations over the domain of the fluid and 
its surface,
respectively.

Note that in the case of infinite domains with 
constant conductivity the electric potential does not appear. 
In this case the integral equation system reduces to 
Eq. (\ref{eq3}) without the
boundary term. A wealth of numerical applications of this 
formulation for infinite
domains can be found in the paper by 
Dobler and R\"adler \cite{DORA}.

In the following sections we will illustrate the general 
approach (\ref{eq3}) and (\ref{eq4}) with  two different 
applications.

\section{Spherical dynamos} 

In this section, $\alpha^2$ dynamos in spherical domains will be considered.
For this case a wealth of quasi-analytical and numerical results are available
from the differential equation approach  which can be used for comparison.

\subsection{The system of radial integral equations}

As usual, we split the magnetic field 
into a poloidal and a toroidal part according to
\begin{eqnarray}
{\mathbf{B}}_{P}&=&{\bf \nabla} \times {\bf \nabla} \times 
\left(\frac{S}{r} \, {\mathbf{r}} 
\right) \; ,\label{eq5}\\
{\mathbf{B}}_{T}&=&{\bf \nabla}  \times \left(\frac{T}{r} \, 
{\mathbf{r}} \right) \; .
\label{eq6}
\end{eqnarray} 
In spherical geometry the defining scalars $S$ and $T$ 
and the electric potential 
are expanded in series of spherical harmonics $Y_{lm}$:
\begin{eqnarray}
S(r,\theta,\phi)&=&\sum_{l,m} s_{lm}(r) Y_{lm}(\theta,\phi)\label{eq7} \; ,\\
T(r,\theta,\phi)&=&\sum_{l,m} t_{lm}(r) Y_{lm}(\theta,\phi)\label{eq8} \; ,\\
\varphi(r,\theta,\phi)&=&\sum_{l,m} 
\varphi_{lm}(r) Y_{lm}(\theta,\phi) \label{eq9}\; .
\end{eqnarray}

For the special case that the only dynamo source is a 
spherically symmetric, isotropic
$\alpha$-effect it was shown  \cite{STE2}
that 
the system of 
integral equations (\ref{eq3}) and (\ref{eq4}) for the magnetic 
field and the electric potential 
can be transformed into the following system of integral equations 
for the expansion coefficients $s_{lm}(r)$ and $t_{lm}(r)$ of the defining scalars:
\begin{eqnarray}
s_{lm}(r)=&\displaystyle \frac{\mu_0 \sigma}{2l+1}&\bigg[
\int\limits_0^r  \frac{{r'}^{l+1}}{r^{l}} \, \alpha(r') \, t_{lm}(r') 
\, dr'+\int\limits_r^R  \frac{r^{l+1}}{{r'}^{l}} \, \alpha(r') \, t_{lm}(r') \, 
dr' \bigg] \; , \label{eq11}
\end{eqnarray}
\begin{eqnarray}
t_{lm}(r)=&\mu_0 \sigma  &\bigg[  \alpha(r) s_{lm}(r) 
- \frac{r^{l+1}}{R^{l+1}} \, \alpha(R) \, s_{lm}(R) \nonumber \\
&&+ \frac{l+1}{2l+1} \, \frac{r^{l+1}}{R^{2l+1}} 
\int\limits_0^R {r'}^l \, \frac{d \alpha(r')}{dr'}
\, s_{lm}(r')  \, dr'\nonumber \\
&&-\frac{l+1}{2l+1} \, \int\limits_0^r  \frac{{r'}^l}{r^l} \, \frac{d 
\alpha(r')}{dr'}
\, s_{lm}(r')  \, dr' \nonumber\\
&&+ \frac{l}{2l+1} \,  \int\limits_r^R \frac{r^{l+1}}{{r'}^{l+1}} \, 
\frac{d \alpha(r')}{dr'}
\, s_{lm}(r')  \, dr' \bigg] \label{eq12} \; .
\end{eqnarray}

This system of integral equations (\ref{eq11}) and (\ref{eq12}) 
is equivalent with the 
system of differential equations and boundary conditions
\begin{eqnarray}
\lambda_l s_{lm}&=&
\frac{1}{\mu_0 \sigma}
\bigg[ \frac{d^2 s_{lm}}{d r^2}-\frac{l(l+1)}{r^2} s_{lm} \bigg]
+\alpha(r) t_{lm}  \; ,
\label{eq11a}\\
\lambda_l t_{lm}&=& \frac{1}{\mu_0 \sigma}
\bigg[\frac{d^2 t_{lm}}{d r^2}- 
\frac{l(l+1)}{r^2} t_{lm}      \bigg]-\frac{d}{dr} \left(\alpha(r)
\frac{d s_{lm}}{dr} \right)\nonumber \\
&&+\frac{l(l+1)}{r^2} \alpha(r)
s_{lm} \; ,
\label{eq12a}\\
t_{lm}(R)&=&R \; \frac{d s_{lm}(r)}{dr} |_{r=R}+{(l+1)} s_{lm}(R)=0 \; ,
\label{eq13a}
\end{eqnarray}
if we set the eigenvalue $\lambda_l$ of the differential equation 
system equal to zero
which corresponds to the steady case.

Note that the effect of the electric potential at 
the boundary 
is already incorporated in Eq. (\ref{eq12}). 
The corresponding term 
ensures that the boundary conditions (\ref{eq13a}) of the 
differential equation system are automatically 
fulfilled in the integral equation method.

\subsection{Numerical implementation} 

In this section, we first present a numerical method for the solution 
of the coupled integral equations (\ref{eq11}) 
and (\ref{eq12}). Then, in order to improve the convergence 
and accuracy further, a convergence accelerating strategy is developed.

\subsubsection{The basic  integral equation solver}

We start with the integral equation system (\ref{eq11}) and (\ref{eq12}).
Setting $x=r/R$, and introducing the notations
\begin{eqnarray}
G_s(x,x')=\left\{ \begin{array}{ll}
\displaystyle\frac{{x'}^{l+1}}{x^l},&0\leq x'< x\\[0.5em]
\displaystyle\frac{x^{l+1}}{{x'}^l},&x<x'\leq 1,
\end{array} \right.  \label{eq13}
\end{eqnarray}
\begin{eqnarray}
G_t(x,x')=\left\{ \begin{array}{ll}
\displaystyle\frac{l}{2l+1}\frac{{x'}^l}{x^l},&0\leq x' < x,\\[0.5em]
\displaystyle\frac{l}{2l+1}\frac{x^{l+1}}{{x'}^{l+1}},& x<x'\leq 1 \; ,
\end{array} \right. 
\label{eq14}
\end{eqnarray}
Eqs. (\ref{eq11}) and (\ref{eq12}) can be  rewritten in the 
following form:
\begin{eqnarray}
s_{lm}(x)=&\displaystyle \frac{\mu_0 \sigma R^2}{2l+1}&\bigg[
\int\limits_0^1 G_s(x,x') \, \alpha(x') \, t_{lm}(x') 
\, dx' \bigg]   \; ,       \label{eq15}
\end{eqnarray}
\begin{eqnarray}
t_{lm}(x)=&\mu_0 \sigma  &\bigg[  \alpha(x) s_{lm}(x) 
- x^{l+1} \, \alpha(1) \, s_{lm}(1) 
\nonumber \\
&&+ \frac{l+1}{2l+1} \, x^{l+1} 
\int\limits_0^1 {x'}^l \, \frac{d \alpha(x')}{dx'}
\, s_{lm}(x')  \, dx'\nonumber \\
&&+\int\limits_0^1 G_t(x,x') \, 
\frac{d \alpha(x')}{dx'}
\, s_{lm}(x')  \, dx' \nonumber\\
&&-\int\limits_0^x \frac{{x'}^l}{x^l} \frac{d \alpha(x')}{dx'}
\, s_{lm}(x')  \, dx' \bigg] \label{eq16}.
\end{eqnarray}

Substituting Eq. (\ref{eq15}) into Eq. (\ref{eq16}) yields
\begin{eqnarray}{\label{eq24}}
w t_{lm}(x)&=&{\tilde{\alpha}}(x)\int\limits_0^1G_s(x,x')
{\tilde{\alpha}}(x')t_{lm}(x')dx' \nonumber\\
&&+\frac{(l+1)}{2l+1} x^{l+1} \int\limits_0^1
\int\limits_0^1{x'}^l
\frac{d{\tilde{\alpha}}(x')}{dx'}
{\tilde{\alpha}}(x'')t_{lm}(x'')
G_s(x',x'')dx''dx' \nonumber\\
&&+\int\limits_0^1\int\limits_0^1 {G}_t(x,x')G_s(x',x''){\tilde{\alpha}}(x'')
t_{lm}(x'')\frac{d{\tilde{\alpha}}(x')}{dx'}dx''dx' \nonumber\\
&&-x^{l+1}{\tilde{\alpha}}(1)
\int\limits_0^1G_s(1,x'){\tilde{\alpha}}(x')t_{lm}(x')dx'\nonumber\\
&&-\int\limits_0^x\int\limits_0^1
\frac{{x'}^l}{x^l}
\frac{d{\tilde{\alpha}}(x')}{dx'}
G_s(x',x''){\tilde{\alpha}}(x'')
t_l(x'')dx'' dx'.
\label{eq17}
\end{eqnarray}
where we use the definition 
\begin{eqnarray}
w=(2l+1)/(\mu_0^2 \sigma^2 C^2 R^2) \; ,
\end{eqnarray}
with $C$ denoting 
a scaling factor of the 
function $\alpha(x)$ according to the new definition 
$\alpha(x)=C {\tilde{\alpha}}(x)$. 
Therefore, the integral equation system 
(\ref{eq15}) and (\ref{eq16}) is reduced to the single integral 
equation (\ref{eq17}). 

For the numerical implementation of Eq. (\ref{eq17}) we 
decided to choose the
classical extended trapezoidal rule. Of course, more 
sophisticated treatments
of the integrals by Gaussian quadratures or Clenshaw-Curtis 
quadrature are as well possible. 
Despite its simplicity, the extended trapezoidal rule
is chosen since it can be easily used as a starting point of 
a convergence accelerating strategy to be discussed in the next
subsection.

Choosing
$N$ equidistant grid points $x_i=i \Delta  x$, with $\Delta x=1/N$,  and 
approximating
the integrals by the extended trapezoidal rule according to
\begin{equation}{\label{eq18}}
\int\limits_0^1f(x)dx\approx \sum\limits_{i=1}^N \frac{1}{2}(f(x_{i-1})+f(x_i)) 
\Delta x \;,
\end{equation}
we obtain for the discretization of Eq. (\ref{eq24}) the following
expression:
\begin{eqnarray}{\label{eq19}}
wt_{lm}(x_i)=&\sum_{j=1}^N& \{ {\tilde{\alpha}}(x_i)c_jG_s(x_i,x_j)
{\tilde{\alpha}}(x_j)\bigtriangleup x  \nonumber\\
&&+\sum_{k=1}^N c_k c_j\frac{d{\tilde{\alpha}}(x_k)}{dx}{\tilde{\alpha}}(x_j)
{G}_t(x_i,x_k)
G_s(x_k,x_j)(\bigtriangleup x)^2\nonumber\\
&&+\frac{l+1}{2l+1}x_i^{l+1}\sum_{k=1}^N c_kc_jx_k^l\frac{d{\tilde{\alpha}}(x_k)}{dx}
{\tilde{\alpha}}(x_j)G_s(x_k,x_j)(\bigtriangleup x)^2\nonumber\\
&&-x_i^{l+1}{\tilde{\alpha}}(1.0)c_jG_s(1.0,x_j){\tilde{\alpha}}(x_j)
\bigtriangleup x\nonumber\\
&&-\sum_{k=1}^{i-1}\frac{x_k^l}{x_i^l}\frac{d{\tilde{\alpha}}(x_k)}
{dx}{\tilde{\alpha}}(x_j)c_j
G_s(x_k,x_j)(\bigtriangleup x)^2\nonumber\\
&& -0.5\frac{d{\tilde{\alpha}}(x_i)}{dx}{\tilde{\alpha}}
(x_j)c_jG_s(x_i,x_j)(\bigtriangleup x)^2 \} t
_{lm}(x_j) \; ,
\end{eqnarray}
where $c_i=1, i=1,2,\cdots, N-1, c_N=0.5$.
Eq. (\ref{eq19}) may be written in the following matrix form:
\begin{equation}{\label{eq25}}
{\mathbf{{A}}}{\mathbf{t}}=w {\mathbf{t}},
\end{equation}
where 
\begin{eqnarray}
{\mathbf{{A}}}=({a}_{ij})_{N\times N} \; ,
\end{eqnarray}
with
\begin{eqnarray}{\label{eq26}}
a_{ij}&=&{\tilde{\alpha}}(x_i)c_jG_s(x_i,x_j)
{\tilde{\alpha}}(x_j)\bigtriangleup x\nonumber\\
&&+\sum_{k=1}^N c_kc_j\frac{d{\tilde{\alpha}}(x_k)}{dx}
{\tilde{\alpha}}(x_j){G}_t(x_i,x_k)
G_s(x_k,x_j)(\bigtriangleup x)^2\nonumber\\
&&+\frac{l+1}{2l+1}x_i^{l+1}\sum_{k=1}^N c_kc_jx_k^l
\frac{d{\tilde{\alpha}}(x_k)}{dx}
{\tilde{\alpha}}(x_j)G_s(x_k,x_j)(\bigtriangleup x)^2\nonumber\\
&&-x_i^{l+1}{\tilde{\alpha}}(1.0)c_jG_s(1.0,x_j)
{\tilde{\alpha}}(x_j)\bigtriangleup x\nonumber\\
&&-\sum_{k=1}^{i-1}\frac{x_k^l}{x_i^l}\frac{d{\tilde{\alpha}}(x_k)}{dx}{\tilde{\alpha}}(x_j)c_j
G_s(x_k,x_j)(\bigtriangleup x)^2\nonumber\\
&&-0.5\frac{d{\tilde{\alpha}}(x_i)}{dx}
{\tilde{\alpha}}(x_j)c_jG_s(x_i,x_j)(\bigtriangleup x)^2  \; .
\end{eqnarray}
This eigenvalue problem can be solved numerically. First, the matrix 
$\mathbf{A}$ is reduced to the Hessenberg form, then the QR algorithm 
can be employed to obtain the eigenvalues $w$ of the matrix ${\mathbf{A}}$ 
and hence those magnitudes $C$ of the functions $\alpha(x)$ for which steady 
dynamos exist. It should be pointed out that, except for the particular 
case $\alpha=const$, the matrix ${\mathbf{{A}}}$ is non-symmetric, hence the
appearance of
complex eigenvalues should be expected.

In the following discussions, the method described 
in this subsection will be called the integral equation solver (IES). 
\subsubsection{Convergence accelerating strategy}

The basic idea of the convergence accelerating strategy 
has been applied in the numerical solutions of 
various integral equations \cite{HACK,CHRI}. This strategy, which is actually 
based 
on the Romberg scheme for the 
numerical quadrature by a extended trapezoidal rule, appears in the 
literature under various notations as  
{\it extrapolation method} \cite{HACK} or 
{\it deferred approach to the limit} \cite{CHRI}. 

In this subsection, this convergence accelerating strategy will be 
adapted for our eigenvalue problem
in order to improve the convergence and accuracy of the 
integral 
equation solver. 

It will be shown in subsection 4.2 that the convergence rate of 
the integral equation solver 
can reach $\sim N^{-2}$. This is also the theoretical 
error estimation of the calculated eigenvalues obtained from the extended 
trapezoidal 
rule with the step size $1/N$ under  the assumption that the kernel is 
sufficiently differentiable \cite{CHRI}. So if $w$ is the exact 
eigenvalue of the integral equation (\ref{eq24}) and $\overline{w}^{(0)}_0$ 
is the eigenvalue calculated by the integral 
equation solver, it is expected \cite{CHRI} that 
\begin{equation}{\label{ad1}}
w=\overline{w}^{(0)}_0+\mu N^{-2}+O(N^{-4}),
\end{equation}
where $\mu$ is a constant. 
In  Eq. (\ref{ad1}), if doubling the grid number, we have
\begin{equation}{\label{ad2}}
w=\overline{w}^{(1)}_0+\frac{1}{4}\mu N^{-2}+O(N^{-4}),
\end{equation}
where $\overline{w}^{(1)}_0$ denotes the eigenvalue calculated by using $2N$ 
grid points.
From Eqs. (\ref{ad1}) and (\ref{ad2}), we obtain
\begin{equation}{\label{ad3}}
w=\overline{w}^{(0)}_1+O(N^{-4}),
\end{equation}
where $\overline{w}^{(0)}_1=(4\overline{w}^{(1)}_0-\overline{w}^{(0)}_0)/3$. 
Therefore, $\overline{w}^{(0)}_1$ approximates the exact eigenvalue $w$ with 
an error which is $O(N^{-4})$. 

The above idea can be extended and a triangular array of entries 
$\overline{w}^{(j)}_k$ is obtained. The array is generated from the 
first column of eigenvalues obtained from discretizing the integrals 
in the integral equation (\ref{eq24}) by the extended trapezoidal rule 
with grid numbers $2^jN, (j=0,1,2,\cdots)$. 
\[
\begin{array}{lllllllll}
j=0&&\overline{w}^{(0)}_0&&&&&&\\
j=1&&\overline{w}^{(1)}_0&&\overline{w}^{(0)}_1&&&&\\
j=2&&\overline{w}^{(2)}_0&&\overline{w}^{(1)}_1&&\overline{w}^{(0)}_2&&\\
j=3&&\overline{w}^{(3)}_0&&\overline{w}^{(2)}_1&&\overline{w}^{(1)}_2&&
\overline{w}^{(0)}_3\\
\vdots&&\vdots&&\vdots&&\vdots&&\vdots\\
\end{array}
\]
The entry $\overline{w}^{(j)}_k$ is placed in the $(j+1)$th position of 
the $(k+1)$th column ($j,k=0,1,2,\cdots$). In general, the entries in 
columns other than the first are obtained by the recurrence relation
\begin{equation}{\label{ad4}}
\overline{w}^{(j)}_k=(4^k\overline{w}^{(j+1)}_{k-1}-
\overline{w}^{(j)}_{k-1})/(4^k-1).
\end{equation}
This idea discussed in this subsection will 
be examined by two examples in Section 4.

\subsection{Numerical examples}

In this section we will treat some example profiles $\alpha(r)$ 
by the developed integral equation solver. In order to validate 
the accuracy of the results they are compared with results known
from other methods.

Note that one has to distinguish between the eigenvalues $\lambda$ 
which appear in the
differential equation system (\ref{eq11a})-(\ref{eq13a}), 
and the 
values  $C$ as they result from the eigenvalues
$w$ of the steady
integral equation system (\ref{eq19}). 
The
eigenvalues $\lambda$ of the differential equation system 
comprise as the real part the growth rate and as the
imaginary part the frequency of the magnetic field mode. 
Both parts have a physical meaning. In contrast to that, 
the values $C$ for the integral equation system give  
critical values for the
intensity of $\alpha$, which are only meaningful 
if they are real. A complex value for $C$ has no physical meaning,
it might only indicate the existence of a
complex eigenvalue $\lambda$ in the vicinity of the 
real part of the 
critical value
$C$.

\subsubsection{Known results}

\begin{figure}[ht]
{\includegraphics[width=12cm]{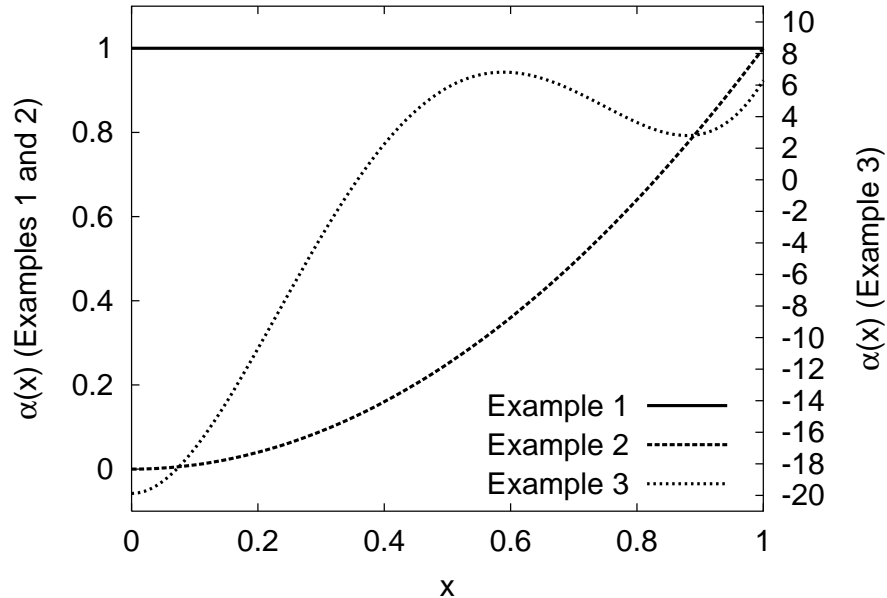}}
\caption{The three considered examples of $\alpha(x)$.
Example 1: $\alpha(x)=C$. Example 2: $\alpha(x)=C x^2$.
Example 3: $\alpha(x)=C (-19.88+347.37 x^2-656.71 x^3+335.52 x^4)$.}
\label{fig1}
\end{figure}

The example profiles that will be considered are the following 
(see Fig. \ref{fig1}): 
\begin{enumerate}
\item $\alpha(x)=C$ ,
\item $\alpha(x)=C  x^2$ ,
\item $\alpha(x)=C (-19.88+
347.37 x^2-656.71 x^3+335.52 x^4)$,
\end{enumerate}
where $C$ denotes the magnitude of the functions.

The first example represents the well-known Krause-Steenbeck dynamo
model, defined by $\alpha(x)=C$. Its eigenvalues $C$ are known to satisfy 
the relation $J_{l+1/2}(C)=0$, with $J_{l+1/2}$ denoting 
the Bessel functions
of degree ${l+1/2}$ \cite{KRRA}. 
In the first row of Table  \ref{eigen1}, we give the first three
eigenvalues $C$ for $l=1$ which we compute by the programme ``Mathematica''.
In order to validate later the results of our integral
equation solver, the results are given with 14 digits after the comma.

\begin{figure}[ht]
{\includegraphics[width=12cm]{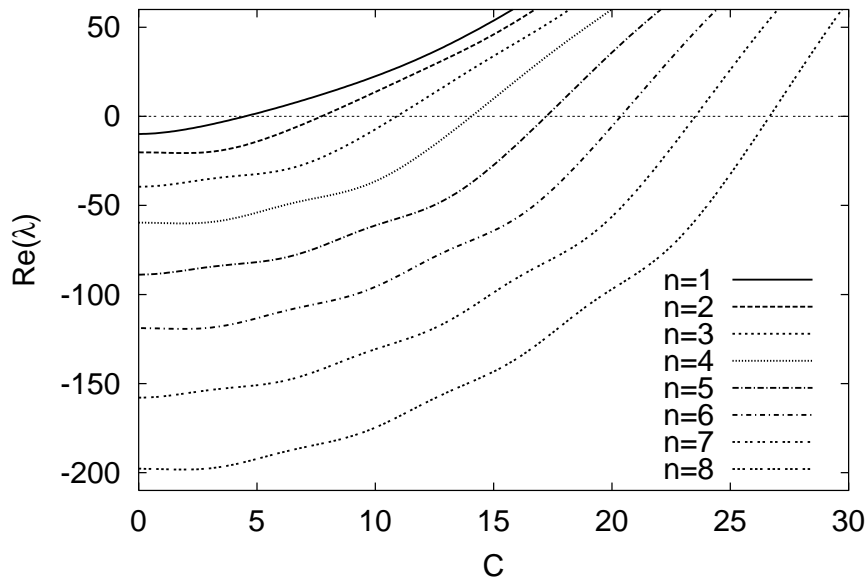}}
\caption{First example: $\alpha(x)=C$. Growth rates 
$Re(\lambda)$ 
for the 
eigenvalues with
$l=1;n=1...8$.}
\label{fig2}
\end{figure}

Fig. \ref{fig2} shows the first
eight eigenvalues $\lambda$ for $l=1$, depending on $C$, 
which are labeled by the radial
wavenumber $n$. These curves result from a differential
equation solver  based on the
shooting technique from Numerical Recipes 
\cite{PTVF}.
For $\alpha(x)=C$ the results of the differential equation solver
code were shown to 
be equivalent 
with the exact value at least until 8 digits after the comma.
For the remaining two examples we believe the accuracy of this 
code to be at least in the same range.
For the Krause-Steenbeck dynamo model, the 
eigenvalues $\lambda$ are 
always real (Fig. \ref{fig2}). 

\begin{figure}[ht]
{\includegraphics[width=12cm]{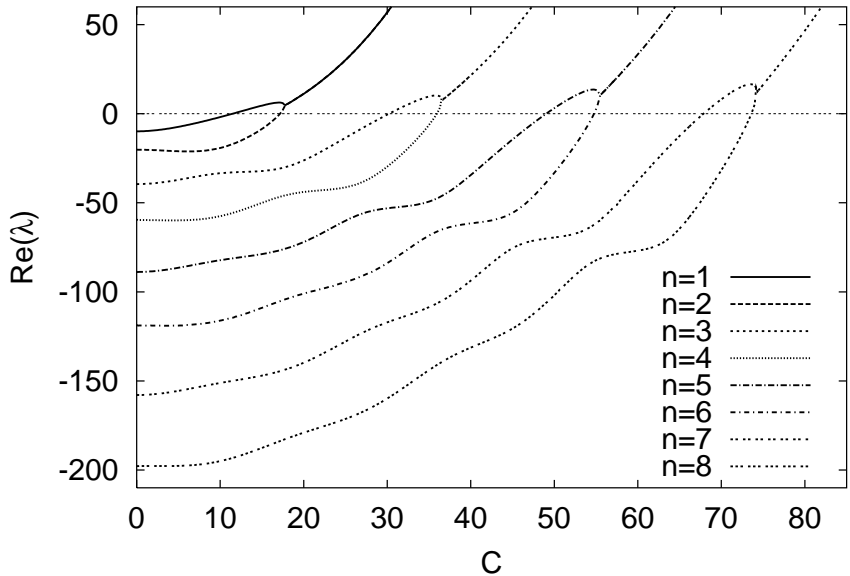}}
\caption{Second example: $\alpha(x)=C  x^2$. 
Growth rates for the 
eigenvalues with
$l=1;n=1...8$. The merging points of two 
neighbouring curves indicate a transition
to a pair of complex conjugated eigenvalues (the imaginary part,
the frequency, is not shown here).}
\label{fig3}
\end{figure}

This situation changes for the second example function, 
$\alpha(x)=C  x^2$. Fig. \ref{fig3} shows again the 
real parts of $\lambda$ for the first
eight eigenvalues for $l=1$. It is clearly visible that the curves
of two neighbouring eigenvalues merge at certain points. At these points, 
two 
real eigenvalues $\lambda$ turn into a pair of complex conjugated 
eigenvalues
(although the frequency is not shown in our plot).
The corresponding critical values of $C$ are shown
in the first row of Table \ref{eigen2}. Here we give only 8 digits 
after the comma which are well justified from the 
accuracy point of view.

\begin{figure}[ht]
{\includegraphics[width=12cm]{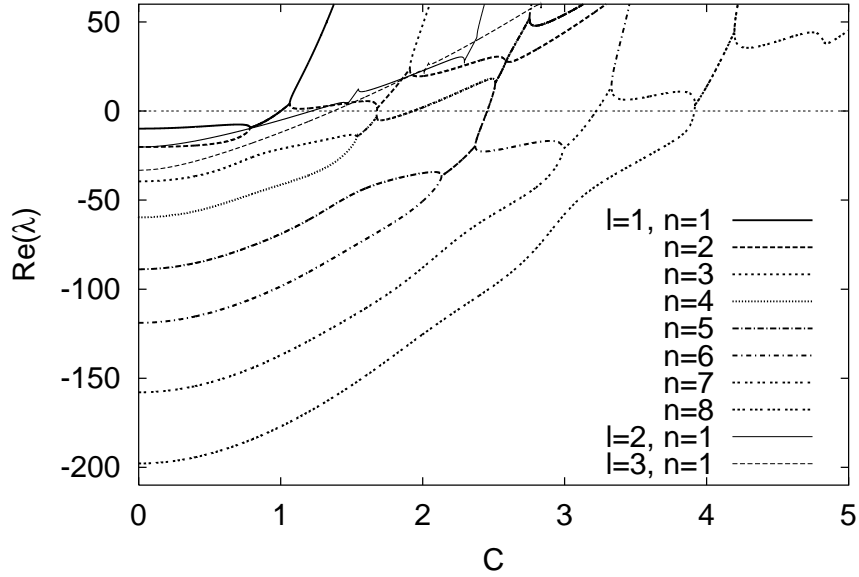}}
\caption{Third example: $\alpha(x)=C  (-19.88+ 347.37 x^2-656.71 x^3+335.52 x^4)$. 
Growth rates for the 
eigenvalues with
$l=1;n=1...8$, and for $l=2;n=1$, $l=3;n=1$. 
At the merging points complex conjugated eigenvalues appear, at the
splitting points two real eigenvalues re-appear. This is a real oscillatory dynamo 
because the eigenmode which 
becomes critical first
has a complex eigenvalue.}
\label{fig4}
\end{figure}

\begin{figure}[ht]
{\includegraphics[width=12cm]{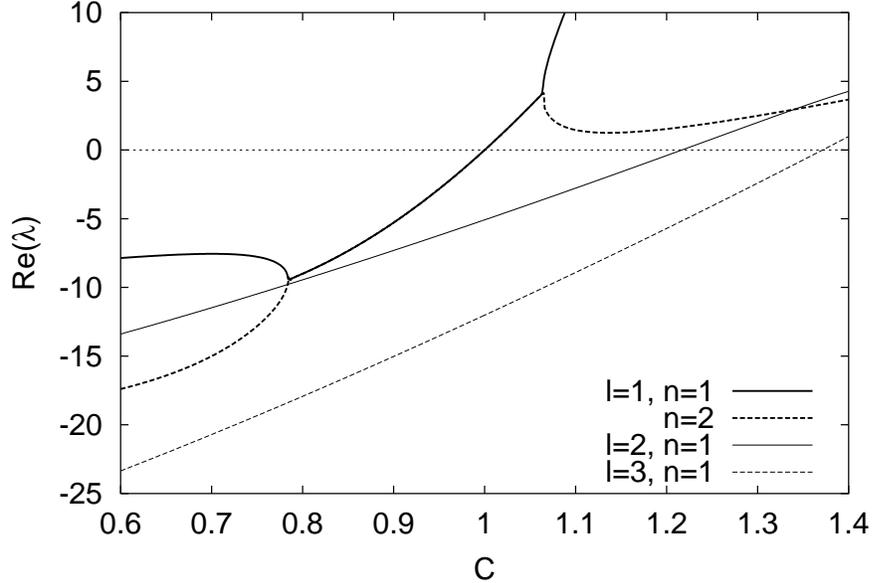}}
\caption{Details of Fig. \ref{fig4}. The mode with $l=1,n=1/2$
crosses zero at $C=0$ where it is oscillatory. All other mode become
critical for higher values of $C$.}
\label{fig5}
\end{figure}

Example 3 has been chosen as it shows a very rich spectral structure,
with complex eigenvalues $\lambda$ at the critical points were 
the growth rates 
are zero. Actually, this function provides the astrophysically interesting 
example of an 
oscillatory $\alpha^2$-dynamo, meaning that the eigenvalue
with zero growth rate is oscillating whereas all the other growth
rates are less than zero \cite{STE3}. 
Fig. \ref{fig4} shows the spectral 
structure,
with its merging and splitting points where two real eigenvalues 
turn into a pair of complex conjugated eigenvalues, and vice versa.
More details close to the critical point can be seen in figure \ref{fig5}.
The corresponding critical values of $C$ are shown
in the first raw of Table \ref{eigen3}. The value in parentheses 
gives the frequency (the imaginary part of $\lambda$) at the critical 
value of $C$.
Here we give only 4 digits after the comma, as we are less interested in
the accuracy problem than in the problem of complex eigenvalues.

\subsubsection{Results of an integral equation solver}

Table \ref{eigen1} shows, for the case of constant $\alpha$, the eigenvalues
of $C$. The first row shows the eigenvalues $C$  resulting 
from a solution of the 
equation $J_{3/2}(x)=0$ by {\it Mathematica}, the remaining 
rows show the results of the integral equation solver (IES)
and of two variants of the accelerated strategy 
for different grid numbers $N$. The first variant, which we call AS1, corresponds 
to the choice $k=1$ in Eq. (\ref{ad4}), the second variant, AS2,
corresponds to $k=2$. The numbers in the first column are the 
number of grid points.

\begin{table}[ht]
\caption{Eigenvalues for $\alpha(x)=C, l=1$.
The first row gives the results of {\it Mathematica}. 
The second
row shows the results obtained by the integral equation solver (IES)
with consecutively doubling the grid number starting from  8. 
The third row represents the results obtained by 
the accelerating strategy one (AS1): 
$(4 \; \overline{w}_0^{(j+1)}-\overline{w}_0^{(j)})/3,j=0,1,\cdots,5$. 
The fourth row represents the results obtained by the 
accelerating strategy two (AS2): $(16 \; \overline{w}_1^{(j+1)}-
\overline{w}_1^{(j)})/15,j=0,1,2,3,4$.}
\label{eigen1}
\begin{center}
\begin{tabular}{cccc}
\hline
&n=1&n=2&n=3\\
\hline
{\it Mathematica}&4.49340945790906&7.72525183693770&10.90412165942889\\
IES&&&\\
8 &4.43504688342757&7.43283337369553&10.09585801749400\\
16&4.47868738573279&7.65070920628469&10.69544817795845\\
32&4.48972062163855&7.70652331733384&10.85151520444580\\
64&4.49248672732462&7.72056385689403&10.89094228165081\\
128&4.49317874264272&7.72407947557310&10.90082507380551\\
256&4.49335177705331&7.72495872368822&10.90329740410711\\
512&4.49339503756767&7.72517855719338&10.90391558878976\\
AS1&&&\\ 
8/16&4.49323421983453&7.72333448381441&10.89531156477993\\
16/32&4.49339836694047&7.72512802101689&10.90353754660825\\
32/64&4.49340876255332&7.72524403674743&10.90408464071915\\
64/128&4.49340941441542&7.72525134846612&10.90411933785708\\
128/256&4.49340945519018&7.72525180639326&10.90412151420764\\
256/512&4.49340945773912&7.72525183502844&10.90412165035065\\
AS2&&&\\      
8/16/32&4.49340931008086&7.72524759016372&10.90408594539681\\
16/32/64&4.49340945559417&7.72525177112947&10.90412111365987\\
32/64/128&4.49340945787289&7.72525183591403&10.90412165099960\\
64/128/256&4.49340945790849&7.72525183692173&10.90412165929768\\
128/256/512&4.49340945790905&7.72525183693745&10.90412165942685\\
\hline
\end{tabular}
\end{center}
\end{table}

\begin{figure}[ht]
{\includegraphics[width=8.4cm]{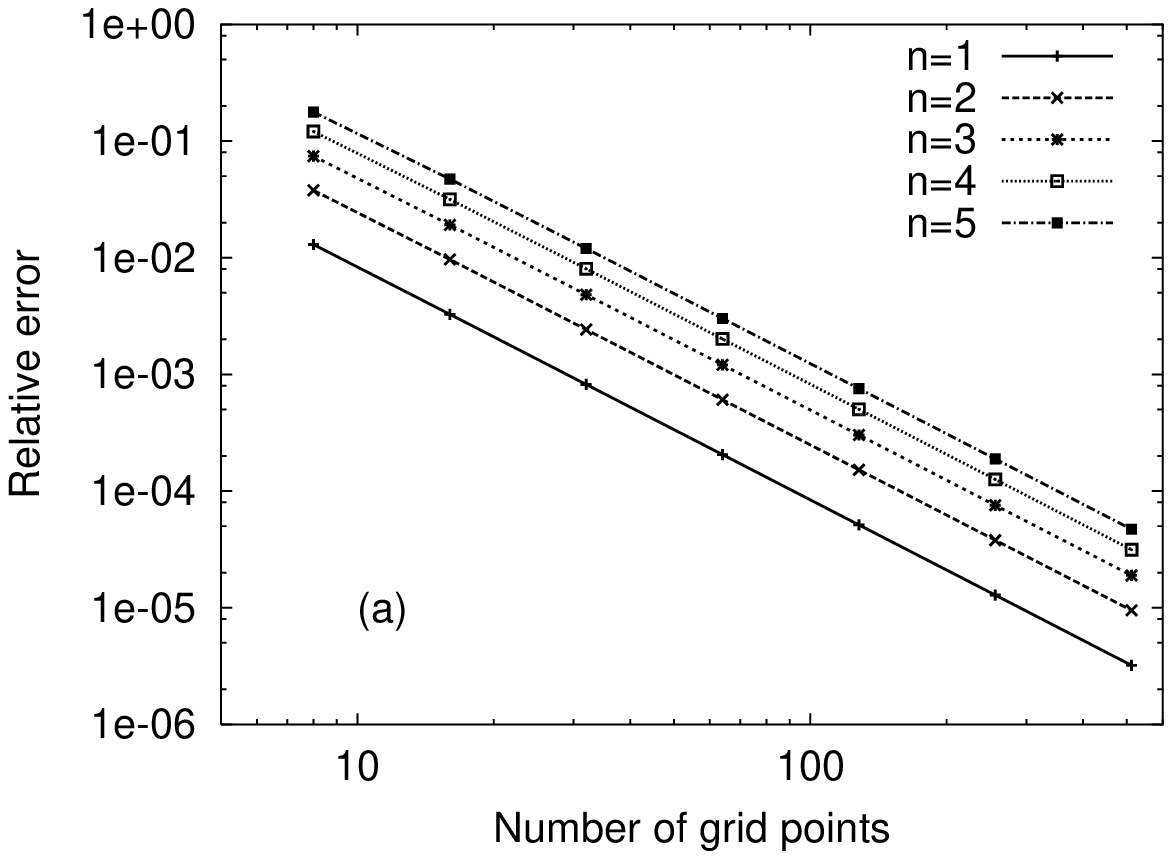}}\\
{\includegraphics[width=8.4cm]{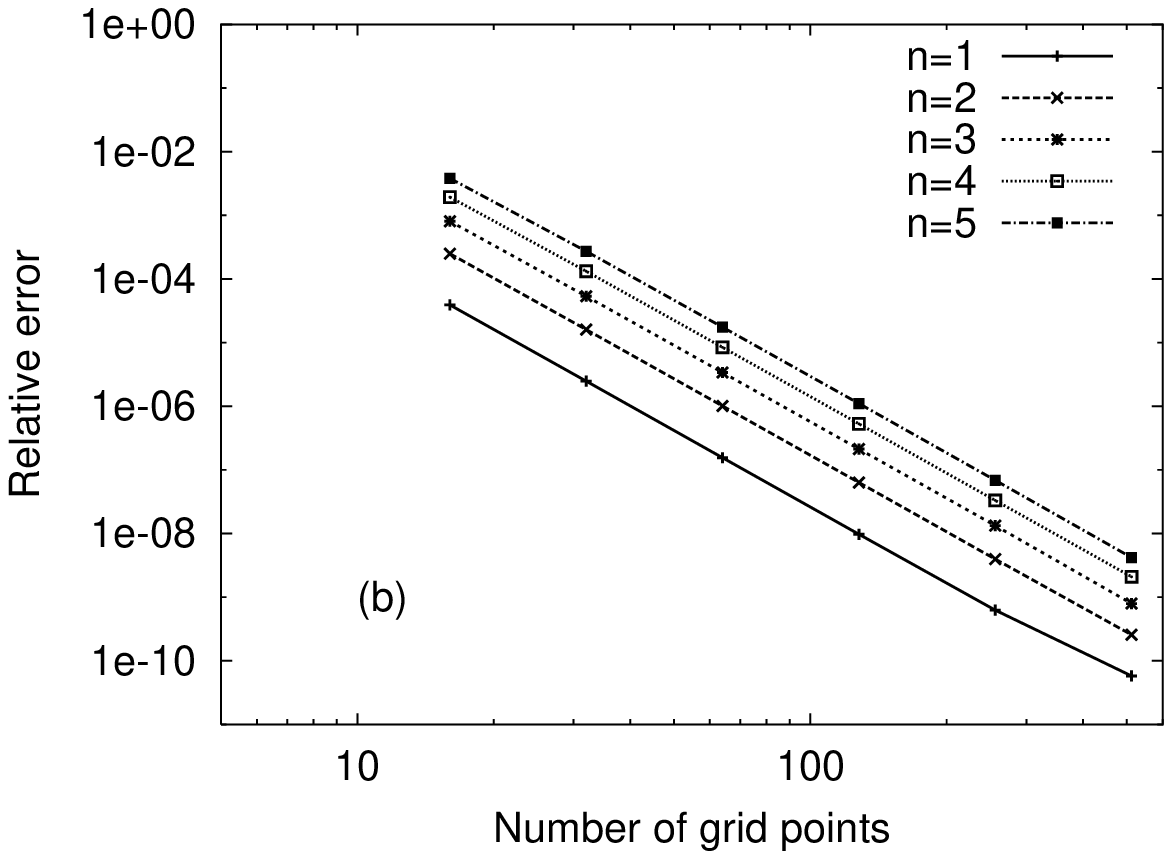}}\\
{\includegraphics[width=8.4cm]{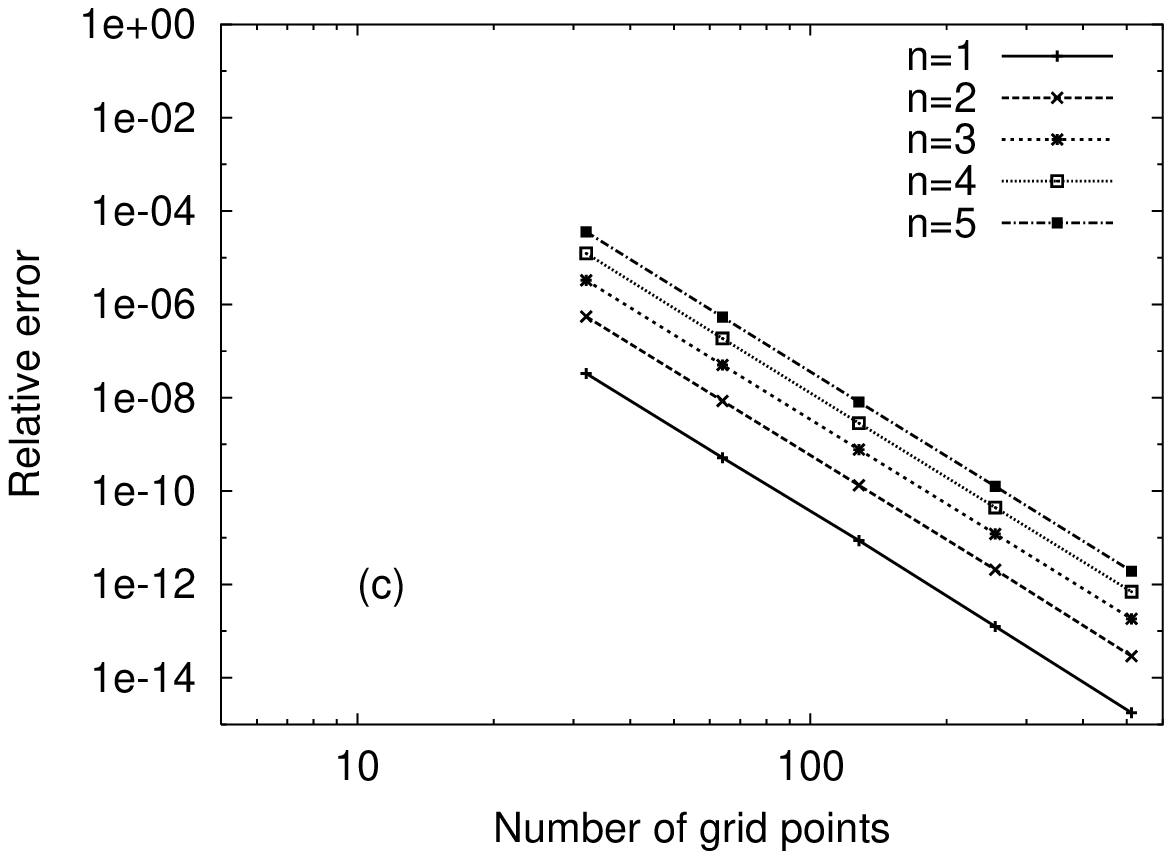}}
\caption{Relative error of the outcome of the integral equation
solver for $\alpha(x)=C$. The comparison is made with results from
{\it Mathematica}. (a) Simple integral equation solver (IES). 
(b) Accelerating strategy 1 (AS1). (c) Accelerating strategy 2 (AS2).}
\label{fig6}
\end{figure}

Based on these values, Fig. \ref{fig6} shows the relative error
of the results for IES, AS1, and AS2. 
For the IES, the error decreases  as 
$\sim N^{-2}$. The error
is larger for the eigenvalues with higher $n$. This is 
no surprise
as the eigenfunctions for higher $n$ are more structured.
For a grid number 512, we can obtain an relative accuracy of the
order 10$^{-5}$. The accuracy can be increased dramatically
if the accelerating strategies AS1 and AS2 are employed. 
For AS1 the convergence is $\sim N^{-4}$, for AS2 it is $\sim N^{-6}$.  
For the latter,
and a grid number 
512,
we obtain a remarkable accuracy between 10$^{-15}$ and 10$^{-12}$. 
Note that, in order to get accuracies better than 10$^{-9}$, it 
was necessary to use the
fourfold precision option of the FORTRAN compiler.

\begin{table}
\caption{Eigenvalues for $\alpha(x)=C  x^2$.
The first row gives the results of the differential equation solver (DES). 
The next
rows show the results obtained by the integral equation solver (IES) by 
consecutively doubling the grid number starting with 8. 
The third row group represents the results obtained by the 
average scheme one (AS1): 
$(4\;\overline{w}_0^{(j+1)}-\overline{w}_0^{(j)})/3,(j=0,1,\cdots,5)$. 
The fourth row group represents the results obtained by the average scheme 
two (AS1): $(16\;\overline{w}_1^{(j+1)}-\overline{w}_1^{(j)})/15,j=0,1,2,3,4$.}
\label{eigen2}
\begin{center}
\begin{tabular}{cccccc}
\hline
&n=1&n=2&n=3&n=4&n=5\\
\hline
DES&11.46714098&17.15742615&30.20482435&35.92762083&49.02331308\\
IES&&&&&\\     
8&10.85648799&16.67863125&26.63454167&42.49377786&79.11841524\\
16&11.30339950&16.96226288&27.81152422&33.41913581&41.42370422\\
32&11.42577229&17.10779976&29.60723780&35.27304970&46.58751516\\
64&11.45677452&17.14499828&30.05610549&35.76666260&48.42948706\\
128&11.46454789&17.15431823&30.16768664&35.88755487&48.87567516\\
256&11.46649262&17.15664913&30.19554253&35.91761522&48.98645261\\
512&11.46697889&17.15723190&30.20250407&35.92512011&49.01410104\\
AS1&&&&&\\
8/16&11.45237001&17.05680675&28.20385174&30.39425513&28.85880054\\
16/32&11.46656322&17.15631206&30.20580899&35.89102100&48.30878547\\
32/64&11.46710860&17.15739779&30.20572806&35.93120024&49.04347770\\
64/128&11.46713901&17.15742488&30.20488035&35.92785229&49.02440453\\
128/256&11.46714086&17.15742609&30.20482783&35.92763533&49.02337843\\
256/512&11.46714098&17.15742616&30.20482458&35.92762175&49.02331718\\
AS2&&&&&\\    
8/16/32&11.46750944&17.16294574&30.33927281&36.25747206&49.60545113\\
16/32/64&11.46714496&17.15747017&30.20572266&35.93387885&49.09245718\\
32/64/128&11.46714104&17.15742668&30.20482384&35.92762909&49.02313298\\
64/128/256&11.46714098&17.15742617&30.20482433&35.92762087&49.02331002\\
128/256/512&11.46714098&17.15742617&30.20482436&35.92762084&49.02331310\\
\hline
\end{tabular}
\end{center}
\end{table}

\begin{figure}
{\includegraphics[width=8.4cm]{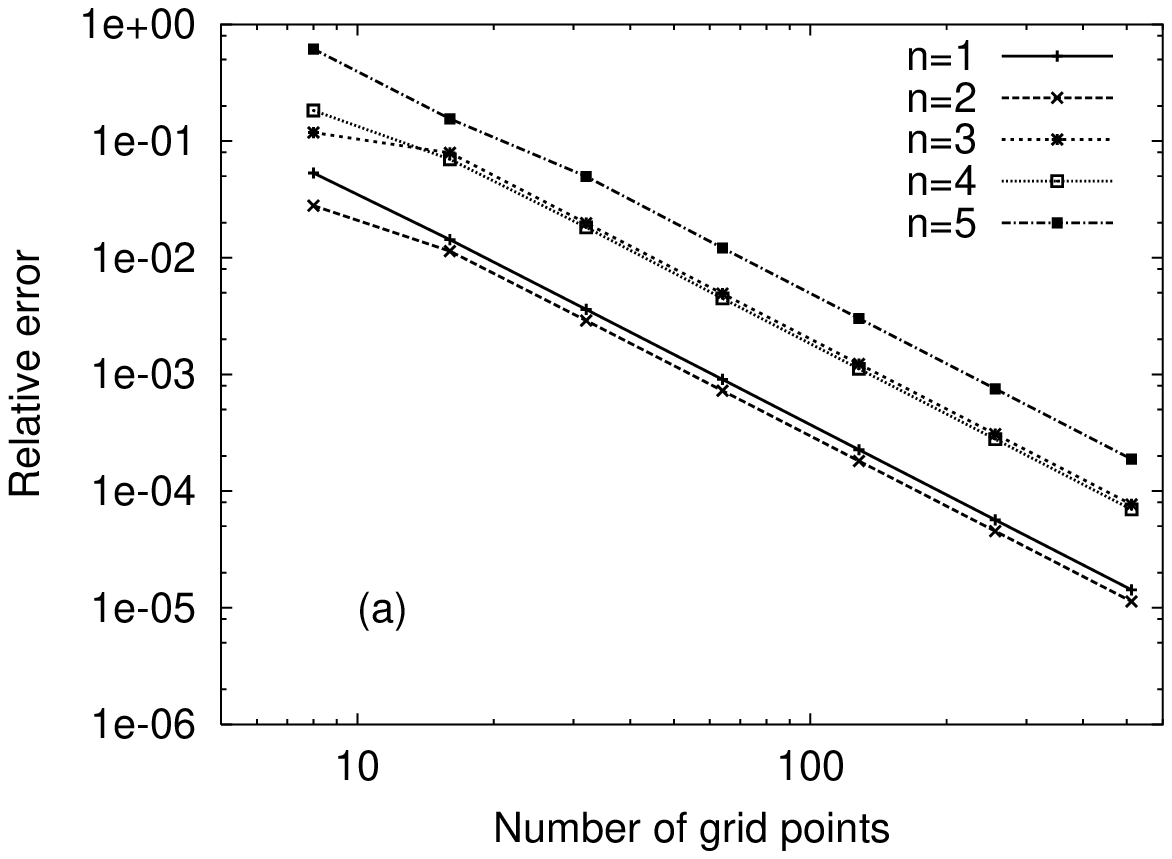}}\\
{\includegraphics[width=8.4cm]{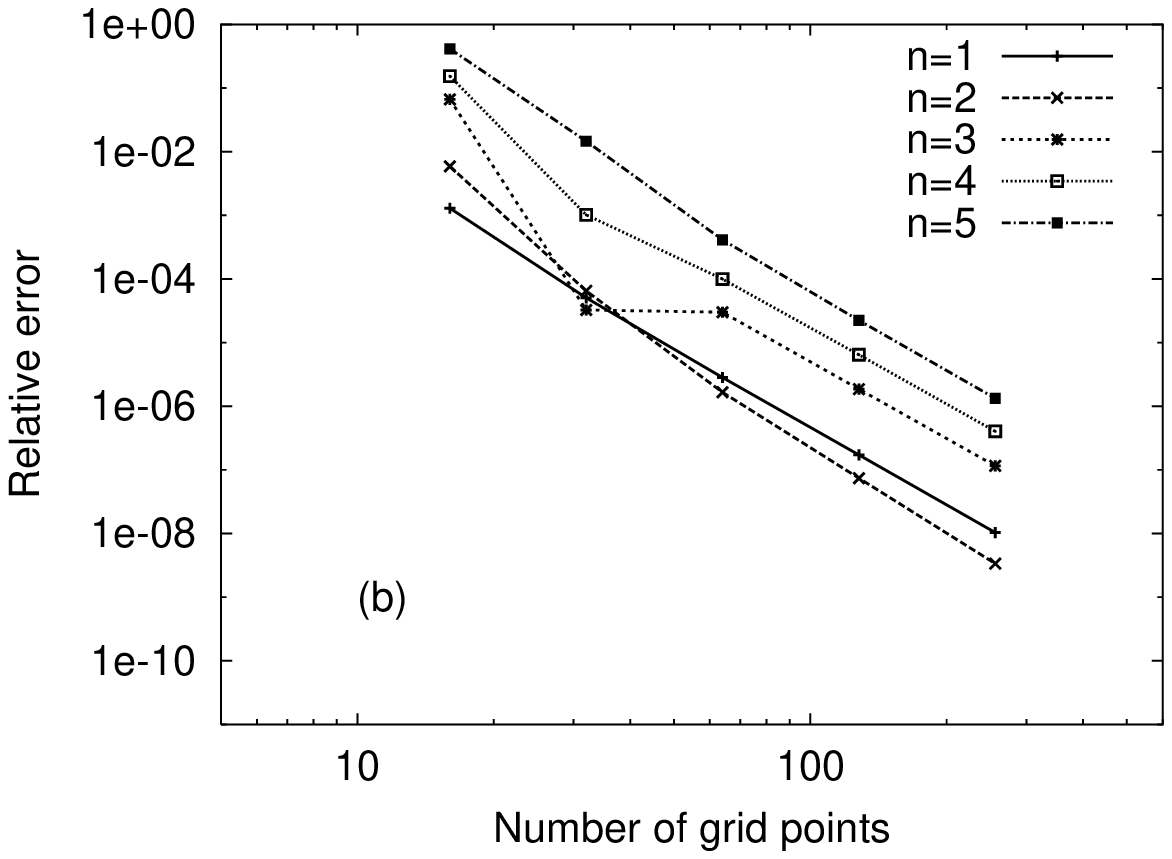}}\\
{\includegraphics[width=8.4cm]{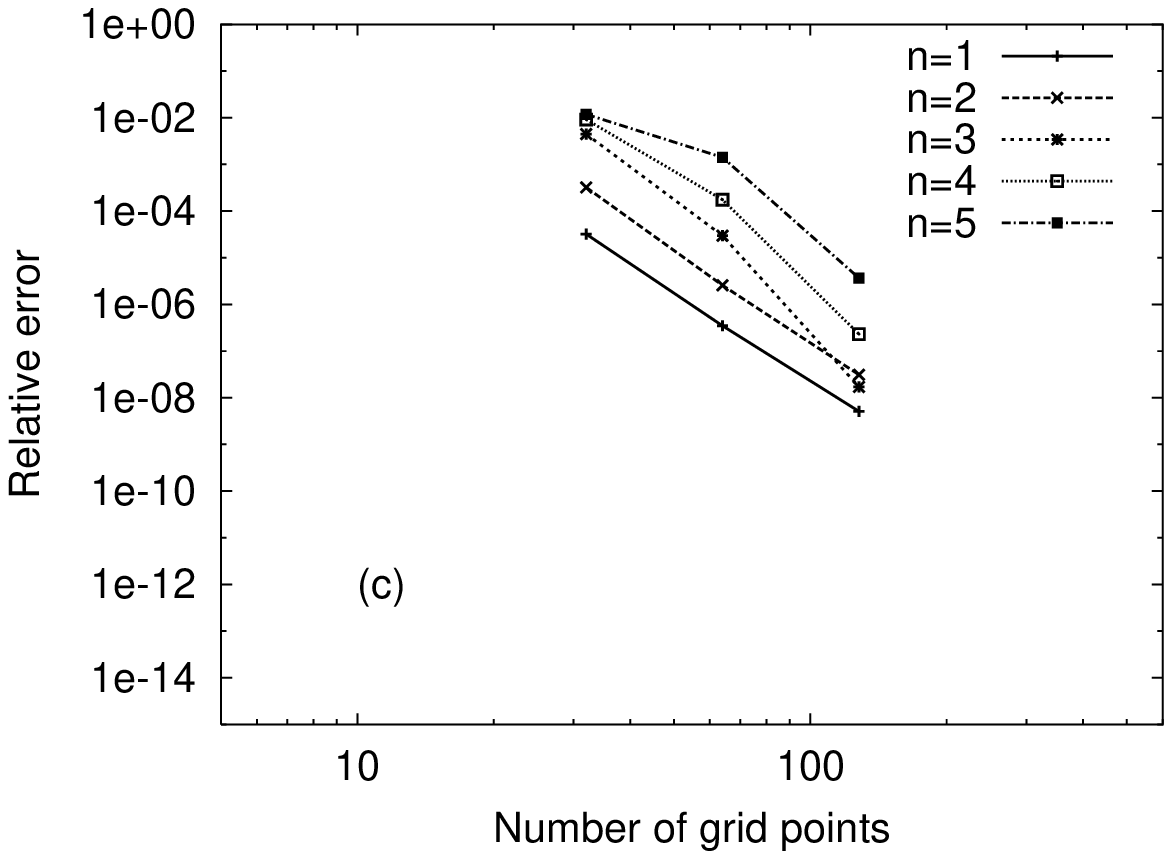}}
\caption{Relative error of the outcome of the integral equation
solver for $\alpha(x)=C x^2$. The comparison is made with results of
a differential equation solver. (a) Simple integral equation solver (IES). 
(b) Accelerating strategy 1 (AS1). (c) Accelerating strategy 2 (AS2).}
\label{fig7}
\end{figure}

Now let us consider the convergence for example 2, $\alpha(x)=C x^2$
(Table \ref{eigen2} and Fig. \ref{fig7}).
The convergence rates for IES, AS1, and AS2 are again 
$\sim N^{-2}$, $\sim N^{-4}$, and $\sim N^{-6}$, respectively.
The errors are typically higher than for the case of constant $\alpha$, which
may have to do with the non-symmetry of the matrix to be inverted.
For AS1 and AS2 we have skipped the last points in Figs.
(\ref{fig7})b and (\ref{fig7})c as we do not have  results from the DES with a 
higher precision.

\begin{table}
\caption{Eigenvalues for $\alpha(x)=C  (-19.88+
347.37 x^2-656.71 x^3+335.52 x^4)$. 
The first row shows the critical value of $C$ resulting from the
differential equation solver (DES). The values in parentheses in the second
row (only for $n=1|2$ and $n=6|7$) give the frequency at this point. 
The remaining rows give the outcomes of the integral equation solver (IES). The 
complex values have no precise physical meaning. 
However it is interesting that the real parts are not very far from the 
correct critical value of $C$.}
\label{eigen3}
\begin{tabular}{cccccc}\hline
&n=1|2&n=3&n=4&n=5&n=6|7\\
\hline
DES&1.0000&1.6788&1.9318&2.4544&3.2223\\
&(4.984)&&&&(13.538)\\
IES&&&&&\\
10&0.9581|1.0695&1.2174&1.8313&3.3357&2.9734$\pm$0.2389 i\\
30&1.1124$\pm$0.1011 i&1.6230&1.9264&2.2805&3.2575$\pm$0.0990 i\\
100&1.1260$\pm$0.1073 i&1.6738&1.9313&2.4384&3.3019$\pm$0.1649 i\\
300&1.1273$\pm$0.1079 i&1.6783&1.9317&2.4526&3.3214$\pm$0.1777 i\\
1000&1.1274$\pm$0.1079 i&1.6788&1.9318&2.4542&3.3237$\pm$0.1790 i\\
3000&1.1274$\pm$0.1079 i&1.6788&1.9318&2.4544&3.3238$\pm$0.1792 i\\
\hline
\end{tabular}
\end{table}

The results for example 3 are represented in Table \ref{eigen3}. 
Below the 
first row that
shows the DES results for the critical value of $C$ we
give for two of the columns the value of the frequency 
that appears at this critical point. For those complex eigenvalues 
$\lambda$ the integral equation method cannot work properly because it is
restricted to the steady, non-oscillatory case. However, it 
is interesting to observe in the rows below 
that the existence of complex eigenvalues
of $\lambda$ is mirrored in the existence of complex eigenvalues of 
$C$. These complex values are unphysical; nevertheless their 
real part is not far from the correct real part, and the imaginary
part indicates oscillatory behaviour.

\section{Matchbox dynamos}
In this section, we consider so-called ''matchbox dynamos'', 
i.e. dynamos in a rectangular box 
which is filled by the electrically conducting fluid and surrounded 
by vacuum. This problem allows us to illustrate how to discretize 
the system of the integral equations ({\ref{eq3}) and (\ref{eq4}) 
in the general case of non-spherical domains.

\subsection{Numerical implementation}
In this subsection, we develop a numerical scheme to 
solve Eqs. (\ref{eq3}) and (\ref{eq4}) directly for the matchbox dynamo. 
In doing so, we have to cope with the singularities of the kernels 
in this integral equation system. Actually, many well developed and 
efficient analytical and numerical methods are available in the 
boundary element method \cite{BREB,PARI} to solve the singular 
integrals, even for integrals with strong singularities. 
For details of the treatment of the singularities in Eqs. (\ref{eq3}) and
(\ref{eq4}), we refer to the appendix.

\subsubsection{Numerical scheme}
In this subsection, the scheme developed in Section 3.1 is 
extended to solve the matchbox dynamo problem making use of the
basic integral 
equations (\ref{eq3}) and (\ref{eq4}). Our scheme 
is very similar to the so-called constant element method which 
is widely applied in the framework of the 
boundary element method \cite{PARI}. 
The difference is that in our scheme the trapezoidal rule 
is utilized for the  discretization of the integrals over elements in order 
to improve the convergence and accuracy. 

Firstly, Eqs. (\ref{eq3}) and (\ref{eq4}) are rewritten in the form
\begin{eqnarray}
B_p({\mathbf{r}})=\frac{\mu_0\sigma}{4\pi}\int_D K_{p,q}
({\mathbf{r}},{\mathbf{r}}')B_q({\mathbf{r}}')dV'-
\frac{\mu_0\sigma}{4\pi}\int_S L_{p}
({\mathbf{r}},{\mathbf{s}}')\varphi({\mathbf{s}}')dS',\label{adeq1}\\
\frac{1}{2}\varphi({\mathbf{s}})=\frac{1}{4\pi}\int_D M_{q}
({\mathbf{s}},{\mathbf{r}}')B_q({\mathbf{r}}')dV'-\frac{1}
{4\pi}\int_Sn_q({\mathbf{s}}')G_q({\mathbf{s}},{\mathbf{s}}')
\varphi({\mathbf{s}}')dS',\label{adeq2}
\end{eqnarray}
where the conventional Einstein's summation has been used. 
$D$ denotes the matchbox, $S$ is the surface of $D$, 
${\mathbf{r}}=(x, y, z)^T$, ${\mathbf{r}}'=(x', y', z')^T$. 
We use the notation
\begin{eqnarray}
K_{p,q}({\mathbf{r}},{\mathbf{r}}')&=&-u_p({\mathbf{r}}')
G_q({\mathbf{r}},{\mathbf{r}}')+u_i({\mathbf{r}}')G_i({\mathbf{r}},
{\mathbf{r}}')\delta_{pq}+\epsilon_{pqi}G_i({\mathbf{r}},{\mathbf{r}}')
\alpha({\mathbf{r}}'),\\
L_{p}({\mathbf{r}},{\mathbf{s}}')&=&\epsilon_{pqi}
n_q({\mathbf{s}}')G_i({\mathbf{r}},{\mathbf{s}}'),\\
M_{q}({\mathbf{s}},{\mathbf{r}}')&=&\epsilon_{piq}u_i({\mathbf{r}}')
G_p({\mathbf{s}},{\mathbf{r}}')+\alpha({\mathbf{r}}') G_q({\mathbf{s}},
{\mathbf{r}}'),
\end{eqnarray}
for $p, q, i=1, 2, 3$, with the definition
\begin{eqnarray}
G_1({\mathbf{r}},{\mathbf{r}}')=\frac{x-x'}{|{\mathbf{r}}-{\mathbf{r}}'|^3},
\; G_2({\mathbf{r}},{\mathbf{r}}')=\frac{y-y'}{|{\mathbf{r}}-{\mathbf{r}}'|^3},\;
G_3({\mathbf{r}},{\mathbf{r}}')=\frac{z-z'}{|{\mathbf{r}}-{\mathbf{r}}'|^3}.
\end{eqnarray} 
As usual, $\delta_{pq}$ denotes the Kronecker symbol, 
and $\epsilon_{pqi}$ is the Levi-Civita symbol.

The matchbox can be expressed as $[0,a]\times[0,b]\times[0,c]$, 
where $a$, $b$ and $c$ are the lengths of the three sides of the 
matchbox. Now we divide this matchbox into $(N-1)^3$ equally sized 
small boxes $D_{ijk} (i, j, k=1, 2, \cdots, N-1)$, which can be 
described as $[x_{i}, x_{i+1}]\times [y_{j}, y_{j+1}]\times [z_{k}, z_{k+1}]$.
The lengths of the intervals $[x_{i}, x_{i+1}]$, $[y_{j}, y_{j+1}]$ 
and $[z_{k}, z_{k+1}]$ are denoted as $\Delta x$, $\Delta y$ and 
$\Delta z$, respectively. The six faces of the matchbox are also 
discretized in a similar manner. For the faces $z=0$ and $z=c$, 
they are divided into the small rectangles $[x_{i}, x_{i+1}]\times 
[y_{j}, y_{j+1}] \; (i, j=1, \cdots, N-1)$; for the faces $y=0$ and $y=b$, 
they are divided into $[x_{i}, x_{i+1}]\times [z_{k}, z_{k+1}] \; (i, k=1, 2, 
\cdots, N-1)$; for the faces $x=0$ and $x=a$, they are 
divided into $[y_{j}, y_{j+1}]
\times[z_{k}, z_{k+1}] \; (j, k=1, 2, \cdots, N-1)$. In the following, we 
denote a representative of these small rectangles as $S_{ij}^{i_s} 
(i_s=1, 2, \cdots, 6)$ which can be expressed as $[x_{1,{i}}^{i_s}, 
x_{1,i+1}^{i_s}]\times [x_{2,{j}}^{i_s}, x_{2,j+1}^{i_s}]$ . The lengths 
of $[x_{1,{i}}^{i_s}, x_{1,i+1}^{i_s}]$ and $[x_{2,{j}}^{i_s}, 
x_{2,j+1}^{i_s}]$ are represented as $\Delta x_1$ and $\Delta x_2$, 
respectively. 

The magnetic fields ''sit'' on the $N^3$ grid points of the volume (including
the surface), whereas the potential ''sits'' on the $6 N^2-12 N +8$ grid 
points of the surface. However, if the grid point is on an edge or a corner, 
we take it
as two or three different grid points by considering it as located on different 
faces. Hence, we have a total of $6 N^2$ electric potential degrees of freedom.

With these definitions, Eqs. (\ref{adeq1}) and (\ref{adeq2}) become
\begin{eqnarray}
B_p({\mathbf{r}})&=&\frac{\mu_0\sigma}{4\pi}\sum_{i'j'k'}\int_{D_{i'j'k'}}
K_{p,q}({\mathbf{r}},{\mathbf{r}}')B_q({\mathbf{r}}')dV'\nonumber \\
&&-
\frac{\mu_0\sigma}{4\pi}\sum_{i_{s1}=1}^6\sum_{i'j'}\int_{S_{i'j'}^{i_{s1}}}
L_{p}({\mathbf{r}},{\mathbf{s}}')\varphi({\mathbf{s}}')dS',
\label{adeq3}\\
\frac{1}{2}\varphi({\mathbf{s}})&=&\frac{1}{4\pi}\sum_{i'j'k'}
\int_{D_{i'j'k'}}M_{q}({\mathbf{s}},{\mathbf{r}}')
B_q({\mathbf{r}}')dV'\nonumber \\&&
-\frac{1}{4\pi}\sum_{i_{s1}=1}^6\sum_{i'j'}
\int_{S_{i'j'}^{i_{s1}}}n_q({\mathbf{s}}')G_q({\mathbf{s}},{\mathbf{s}}')
\varphi({\mathbf{s}}')dS'.\label{adeq4}
\end{eqnarray}
For the integrals over $D_{i'j'k'}$, the application of the  
trapezoidal rule leads to
\begin{eqnarray}
\sum_{i'j'k'}\int_{D_{i'j'k'}} K_{p,q}({\mathbf{r}},
{\mathbf{r}}')B_q({\mathbf{r}}')dV'\approx\nonumber\\
\sum_{i'j'k'}c_{i'} c_{j'} c_{k'} K_{p,q}
({\mathbf{r}},{\mathbf{r}}_{i'j'k'})B_q({\mathbf{r}}_{i'j'k'})
\Delta x\Delta y\Delta z,\label{adeq5}\\
\sum_{i'j'k'}\int_{D_{i'j'k'}} M_{q}({\mathbf{s}},{\mathbf{r}}')
B_q({\mathbf{r}}')dV'\approx\nonumber \\ 
\sum_{i'j'k'}c_{i'} c_{j'} c_{k'}
M_{q}({\mathbf{s}},{\mathbf{r}}_{i'j'k'})
B_q({\mathbf{r}}_{i'j'k'})\Delta x\Delta y\Delta z,\label{adeq6}
\end{eqnarray}
where $c_{i'}$ is defined as $c_1=0.5, \; c_N=0.5, \; c_{i'}=1.0, 
\; i'=2, 3, \cdots, N-1$, ${\mathbf{r}}_{i'j'k'}=(x_{i'}, y_{j'}, x_{k'})^T$.
Similarly, for the integrals over $S_{i'j'}^{i_{s1}}$, we have
\begin{eqnarray}
\sum_{i_{s1}=1}^6\sum_{i'j'}\int_{S_{i'j'}^{i_{s1}}}
L_{p}({\mathbf{r}},{\mathbf{s}}')
\varphi({\mathbf{s}}')dS'=\nonumber \\
\sum_{i_{s1}=1}^6
\sum_{i'j'}c_{i'} c_{j'} L_{p}({\mathbf{r}},
{\mathbf{s}}_{i'j'}^{i_{s1}})
\varphi({\mathbf{s}}_{i'j'}^{i_{s1}})\Delta x_1\Delta x_2,\label{adeq7}\\
\sum_{i_{s1}=1}^6\sum_{i'j'}\int_{S_{i'j'}^{i_{s1}}}n_q({\mathbf{s}}')
G_q({\mathbf{s}},{\mathbf{s}}')
\varphi({\mathbf{s}}')dS'=\nonumber \\ 
\sum_{i_{s1}=1}^6\sum_{i'j'}c_{i'} c_{j'}
n_q({\mathbf{s}}_{i'j'}^{i_{s1}})G_q({\mathbf{s}},{\mathbf{s}}_{i'j'}^{i_{s1}})
\varphi({\mathbf{s}}_{i'j'}^{i_{s1}}) \Delta x_1\Delta x_2.\label{adeq8}
\end{eqnarray}
Substituting Eqs. (\ref{adeq5}-\ref{adeq8}) into Eqs. (\ref{adeq3}) 
and (\ref{adeq4}) and letting ${\mathbf{r}}={\mathbf{r}}_{ijk}, 
{\mathbf{s}}={\mathbf{s}}_{ij}^{i_s}$, we obtain
\begin{eqnarray}
B_p({\mathbf{r}}_{ijk})&=&\frac{\mu_0\sigma}{4\pi}\sum_{i'j'k'}c_{i'}c_{j'}
c_{k'} K_{p,q}({\mathbf{r}}_{ijk},{\mathbf{r}}_{i'j'k'})
B_q({\mathbf{r}}_{i'j'k'})\Delta x\Delta y\Delta z- \nonumber\\
&&\frac{\mu_0\sigma}{4\pi}\sum_{i_{s1}=1}^6\sum_{i'j'}c_{i'}c_{j'}
L_{p}({\mathbf{r}}_{ijk},{\mathbf{s}}_{i'j'}^{i_{s1}})
\varphi({\mathbf{s}}_{i'j'}^{i_{s1}})\Delta x_1\Delta x_2,\label{adeq9}\\
\frac{1}{2}\varphi({\mathbf{s}}_{ij}^{i_s})&=&\frac{1}{4\pi}
\sum_{i'j'k'}c_{i'} c_{j'} c_{k'} M_{q}({\mathbf{s}}_{ij}^{i_s},
{\mathbf{r}}_{i'j'k'})B_q({\mathbf{r}}_{i'j'k'})\Delta x\Delta y\Delta z-\nonumber\\
&&\frac{1}{4\pi}\sum_{i_{s1}=1}^6\sum_{i'j'} c_{i'} c_{j'}
n_q({\mathbf{s}}_{i'j'}^{i_{s1}})G_q({\mathbf{s}}_{ij}^{i_s},
{\mathbf{s}}_{i'j'}^{i_{s1}})\varphi({\mathbf{s}}_{i'j'}^{i_{s1}})
\Delta x_1\Delta x_2,\label{adeq10}
\end{eqnarray}
where $i,j,k,i',j',k'=1, 2, \cdots, N$, and $i_s, i_{s1}=1, 2, \cdots, 6$. 

Note that when ${\mathbf{r}}_{ijk}$ belongs to $D_{i'j'k'}$, 
a weak singularity of the first integral of the right hand side 
of Eq. (\ref{adeq3}) occurs. We can employ the strategy 
discussed in the appendix to deal with such a singularity. 
For example, we can eliminate a small box $[x_{i+1}-\frac{1}{8}
\Delta x, x_{i+1}]\times [y_{j+1}-\frac{1}{8}\Delta y, y_{j+1}]
\times [z_{k+1}-\frac{1}{8}\Delta z, z_{k+1}]$ from $D_{ijk}$ 
when dealing with the weak singularity caused by 
setting ${\bf{r}}$ to ${\bf{r}}_{ijk}$ in the integral 
\[
\int_{D_{ijk}} K_{p,q}({\mathbf{r}},{\mathbf{r}}')B_q({\mathbf{r}}')dV' . 
\]
The overall effect of doing so is equivalent to setting 
$K_{p,q}({\mathbf{r}}_{ijk},{\mathbf{r}}_{i'j'k'})$ to 
zero in Eq. (\ref{adeq9}) when ${\mathbf{r}}_{ijk}={\mathbf{r}}_{i'j'k'}$. 

A similar technique can be applied to handle the 
singularity appearing in the second integral of the 
right hand side of Eq. (\ref{adeq4}). For example, we consider 
the following integral
\[
\int_{S_{ij}^{i_s}}n_q({\mathbf{s}}')G_q({\mathbf{s}}_{ij}^{i_s},
{\mathbf{s}}')\varphi({\mathbf{s}}')dS'.
\]
Since the point ${\mathbf{s}}_{ij}^{i_s}$ belongs to $S_{ij}^{i_s}$, 
it results in a singularity of this integral. We can define 
a small piece of surface as $S_{\epsilon ij}^{i_s}=[x_{1,i+1}-
\frac{1}{4}\Delta x_1, x_{1,i+1}]\times [x_{2,j+1}-\frac{1}{4}\Delta x_2, x_{2,j+1}]$. 
When proceeding the discretization, we just replace 
$S_{ij}^{i_s}$ by $S_{ij}^{i_s}-S_{\epsilon ij}^{i_s}$ and neglect the 
small piece $S_{\epsilon ij}^{i_s}$. This is also equivalent to 
setting $G_q({\mathbf{s}}_{ij}^{i_s},{\mathbf{s}}_{i'j'}^{i_{s1}})$ 
to zero when ${\mathbf{s}}_{ij}^{i_s}={\mathbf{s}}_{i'j'}^{i_{s1}}$. 

Note that similar procedures can be employed to avoid the singularities of 
the other integrals in Eqs. (\ref{eq3}) and (\ref{eq4}). 

Eqs. (\ref{adeq9}) and (\ref{adeq10}) can be rewritten in the matrix form
\begin{eqnarray}
{\mathbf{X}}_B&=&\mu_0\sigma ({\mathbf{E}}
{\mathbf{X}}_B-{\mathbf{D}}
{\mathbf{X}}_\varphi),\label{adeq11}\\
0.5 \; {\mathbf{X}}_\varphi&=&{\mathbf{H}}
{\mathbf{X}}_B-{\mathbf{A}}{\mathbf{X}}_\varphi,\label{adeq12}
\end{eqnarray}
where
\begin{eqnarray}
{\mathbf{X}}_B&=&(B_1({\mathbf{r}}_{111}),B_2({\mathbf{r}}_{111}), 
\cdots,
B_2({\mathbf{r}}_{NNN}),B_3({\mathbf{r}}_{NNN}))^T,\\
{\mathbf{X}}_\varphi&=&(\varphi({\mathbf{s}}_{11}^{1}), 
\varphi({\mathbf{s}}_{12}^{1}), \cdots, \varphi({\mathbf{s}}_{NN}^{6}))^T,
\end{eqnarray}
\begin{eqnarray}
&&A((i_s-1)N^2+(i-1)N+j,(i_{s1}-1)N^2+(i'-1)N+j')\nonumber\\
&=&\frac{1}{4\pi} \; c_{i'} \; c_{j'} \; 
n_q({\mathbf{s}}_{i'j'}^{i_{s1}}) \;
G_q({\mathbf{s}}_{ij}^{i_s},{\mathbf{s}}_{i'j'}^{i_{s1}}) \; \Delta x_1 
\; \Delta x_2,
\end{eqnarray}
\begin{eqnarray}
&&H((i_s-1)N^2+(i-1)N+j,3N^2(i'-1)+3N(j'-1)+3(k'-1)+q)\nonumber\\
&=&\frac{1}{4\pi} \; c_{i'} \; c_{j'} \; c_{k'} \; 
M_{q}({\mathbf{s}}_{ij}^{i_s},{\mathbf{r}}_{i'j'k'}) \; 
\Delta x \; \Delta y \; \Delta z,
\end{eqnarray}
\begin{eqnarray}
&&E(3N^2(i-1)+3N(j-1)+3(k-1)+p,\nonumber\\ 
&&\;\;\;\;\;3N^2(i'-1)+3N(j'-1)+3(k'-1)+q)\nonumber\\
&=&\frac{1}{4\pi} \; c_{i'} \; c_{j'} \; c_{k'} \; 
K_{p,q}({\mathbf{r}}_{ijk},{\mathbf{r}}'_{i'j'k'}) \;
\Delta x \; \Delta y \; \Delta z,
\end{eqnarray}
\begin{eqnarray}
&&D(3N^2(i-1)+3N(j-1)+3(k-1)+p,(i_{s1}-1)N^2+(i'-1)N+j')\nonumber\\
&=&\frac{1}{4\pi} \; c_{i'} \; c_{j'} \; L_{p}
({\mathbf{r}}_{ijk}, {\mathbf{s}}_{i'j'}^{i_{s1}}) \; \Delta x_1 \; 
\Delta x_2,
\end{eqnarray}
with $p,q=1, 2, 3$, $i, j, k, i', j', k'=1, 2, \cdots, N$ and 
$i_s=1, 2, \cdots, 6$.

From Eq. (\ref{adeq12}), we obtain
\begin{eqnarray}{\label{adeq13}}
{\mathbf{X}}_\varphi=(0.5 \; {\mathbf{I}}+{\mathbf{A}})^{-1}
{\mathbf{H}}{\mathbf{X}}_B.
\end{eqnarray}
For the inversion of the matrix $0.5\; {\mathbf{I}}+{\mathbf{A}}$ 
some particular care is needed. 
Physically, the electric potential is defined only up to an additive constant, 
which implies that the matrix $0.5\; {\mathbf{I}}+{\mathbf{A}}$ 
is singular. This difficulty can be removed by 
applying the deflation method \cite{BARN}, 
which is widely used, e.g., in the context of electro- and magnetoencephalography
\cite{HAEM}.
Actually, the matrix ${\mathbf{A}}$ 
can be replaced by
\begin{eqnarray}{\label{adeq14}}
{\mathbf{A}}_1={\mathbf{A}}+\frac{1}{6N^2}{\mathbf{I}}_1,
\end{eqnarray}
where we denote by ${\mathbf{I}}_1$ a quadratic matrix of the order
$6 N^2\times 6N^2$ whose entries are all equal to one. 
Thus, Eq. (\ref{adeq13}) becomes
\begin{eqnarray}{\label{adeq15}}
{\mathbf{X}}_\varphi=(0.5\;{\mathbf{I}}+{\mathbf{A}}_1)^{-1}{\mathbf{H}}
{\mathbf{X}}_B.
\end{eqnarray}
Substituting Eq. (\ref{adeq15}) into Eq. (\ref{adeq11}) yields
\begin{eqnarray}{\label{adeq16}}
\frac{1}{R_m}{\mathbf{X}}_B=({\mathbf{E}}-{\mathbf{D}}(0.5\;{\mathbf{I}}+
{\mathbf{A}}_1)^{-1}{\mathbf{H}}){\mathbf{X}}_B,
\end{eqnarray}
where $R_m$ is the magnetic Reynolds number. This eigenvalue problem 
can be solved by the QR method, which gives the critical magnetic 
Reynolds numbers and the corresponding modes of the magnetic 
field and the electric potential. Although this numerical scheme 
is presented for the dynamo action in the matchbox, it can be 
easily extended to solve steady dynamo problems in other domains.
\subsection{Numerical examples}

In the following we will treat $\alpha^2$ dynamos with a constant value 
$\alpha=C$ 
within
a rectangular box.
The most interesting situation is with vacuum in the
exteriour of the box. Only for 
the cubic box we consider also the case with 
the exteriour space  having the same conductivity as the interiour, in order to
compare the
results with the analytically known ones for spheres of comparable sizes.

\begin{table}[ht]
\caption{The first eigenvalue for a dynamo with $\alpha({\bf r})=C$ within 
a cubic box of
sidelength 2 for the cases of conducting and insulating exteriour. 
The first two rows give, for the sake of comparison, the analytically known
critical values for an enclosed sphere with
radius 1, and for an enclosing sphere with radius $\sqrt{3}$. 
The remaining rows show the numerical results of the integral equation approach
for different grid point numbers $N$ in one direction.
Note that there is a threefold degeneracy of the first eigenvalue due
to the symmetry of the problem.}
\label{eigen4}
\begin{tabular}{ccc}\hline
&Conducting&Vacuum\\
\hline
$C_{sphere}$&3.506&4.493\\
$C_{sphere}/\sqrt{3}$&2.431&3.116\\
IES&&\\
5&3.524&4.254\\
6&3.292&3.996\\
7&3.170&3.866\\
8&3.098&3.793\\
9&3.052&3.750\\
10&3.021&3.723\\
12&2.982&3.694\\
15&2.952&3.678\\  
\hline
\end{tabular}
\end{table}

In Table \ref{eigen4}, we show for the cubic box 
the first eigenvalues $C$ in dependence
on the grid number $N$ in one direction (the total grid number is then $N^3$). 
To the best of our knowledge, there are no values available in the literature
to compare our results with. However, there is at least a plausibility check 
for our results.
Imagine two spheres, the first one, with radius 1, being embedded 
neatly into our cubic box, the
second one, with radius $\sqrt{3}$ enclosing the box. It should be expected that 
the eigenvalues for the cubic box are between those for the two 
spheres. As can be seen from Table \ref{eigen4}, this 
is indeed the case, both for the case of a conducting outer space and
for an insulating exteriour.

\begin{figure}[ht]
{\includegraphics[width=8.4cm]{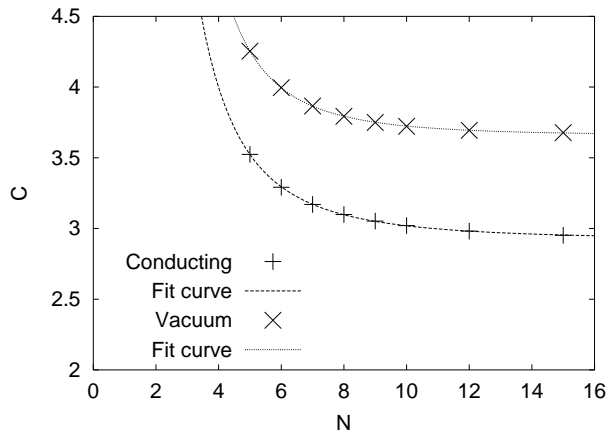}}\\
\caption{Convergence of the three-fold degenerated lowest eigenvalue for the cubic box 
with increasing grid points $N$ 
for the cases of conducting and vacuum exteriour.
The fit curves
are $C(N)=2.919+39.544 N^{-2.599}$  for the case of conducting exteriour and
$C(N)=3.656+91.301 N^{-3.124}$ for the insulating exteriour case, respectively. 
$N$ is the number
of grid points in one direction, hence the total number of grid points is $N^3$.}
\label{fig8}
\end{figure}

The convergence of the eigenvalue for increasing $N$ is 
illustrated in Fig. \ref{fig8}. We 
have made a fit of the eigenvalue data
to the free parameters $f$, $g$, and $h$ in the function $C(N)=f+g  N^{3 h}$. 
The parameter $h$ gives 
the convergence rate for increasing
$N$, whereas the parameter $f$ gives a reasonable estimate of the
true eigenvalue. For the case of conducting exteriour this value is 
2.919 as compared with 3.506 for the enclosed sphere and 
2.431 for the enclosing sphere. For the case of vacuum  the value is 
3.656 as compared with 4.493 for the enclosed sphere and 
3.116 for the enclosing sphere. Hence, the results are plausible in both cases.

As for the convergence rate, the value -1.041 indicates a faster convergence 
than $\log \tilde{N}/\tilde{N}$ which was found 
by Dobler and 
R\"adler \cite{DORA} (we use $\tilde{N}$ for their 
total number of grid points in order
to distinguish it from
our number $N$ of grid points in one direction). 
This better convergence rate should be attributed to the
use of the trapezoidal rule
for the integration instead of the constant element method, what we  have also
confirmed by comparative computations with the latter method.

\begin{table}
\caption{The three first eigenvalues for $\alpha^2$ dynamos in a rectangular box, 
in dependence on the ratios
of side lenghts.}
\label{eigen5}
\begin{tabular}{ccccc}\hline
N&Side lengths&1. EV&2. EV&3. EV\\
\hline
8&1.0:1.0:1.0& 3.793 &3.793 &3.793\\
8& 1.0:1.0:0.8& 4.072 &4.128 &4.128\\
8& 1.0:1.0:0.6& 4.524 &4.674 &4.674\\
8& 1.0:1.0:0.4& 5.322 &5.515 &5.515\\
8& 1.0:0.8:0.6& 4.878 &4.956 &4.500\\
11& 1.0:0.8:0.6& 4.728 &4.898 &4.934\\
8& 1.0:0.8:0.5& 5.235 &5.350 &5.436\\
\hline
\end{tabular}
\end{table}

If we replace  the cube by rectangular boxes with different ratios of the
side lengths we can see in Table \ref{eigen5}
that the three-fold degeneration of the eigenvalues is lifted.

\begin{figure}
{\includegraphics[width=7cm]{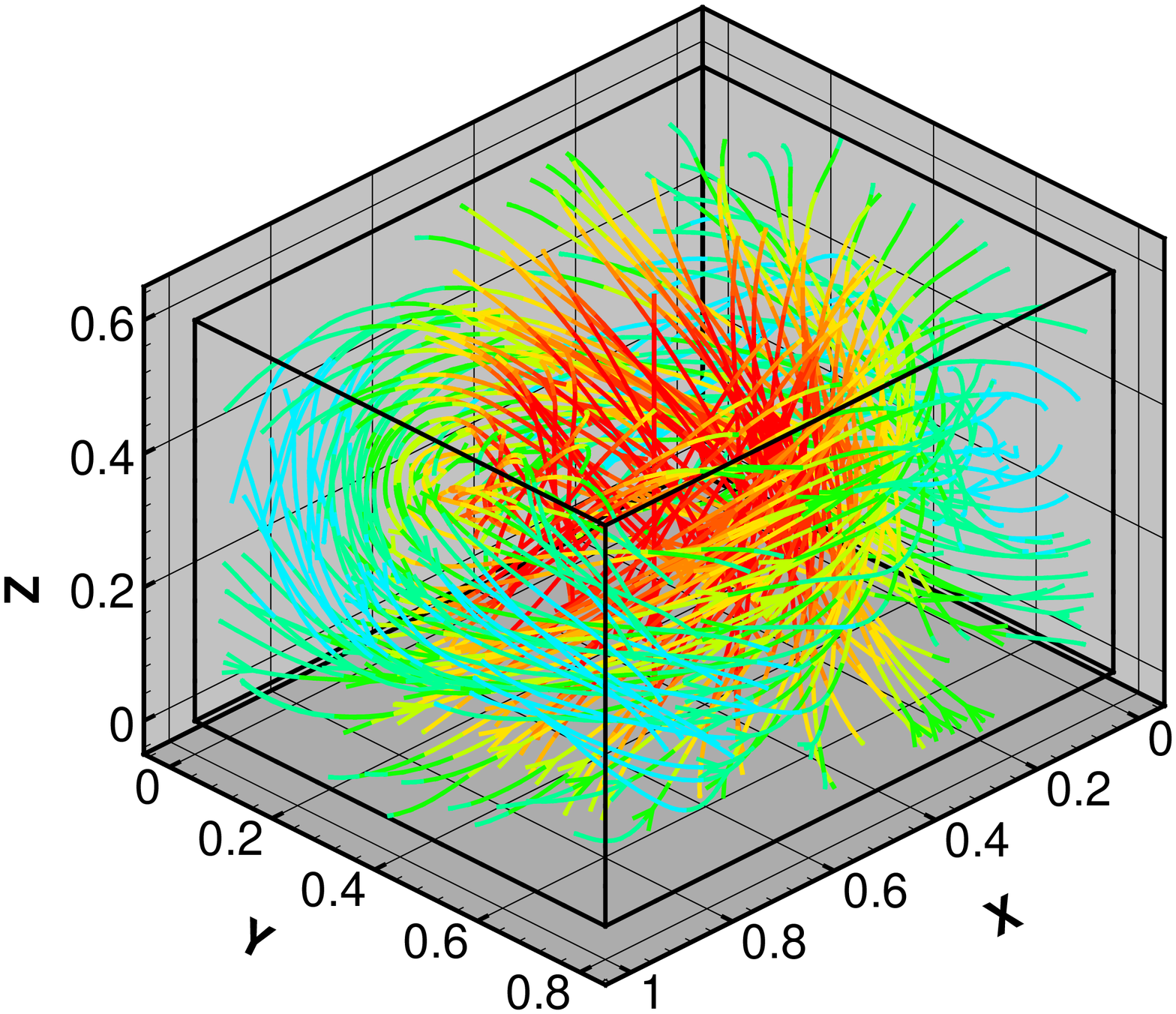}}
{\includegraphics[width=7cm]{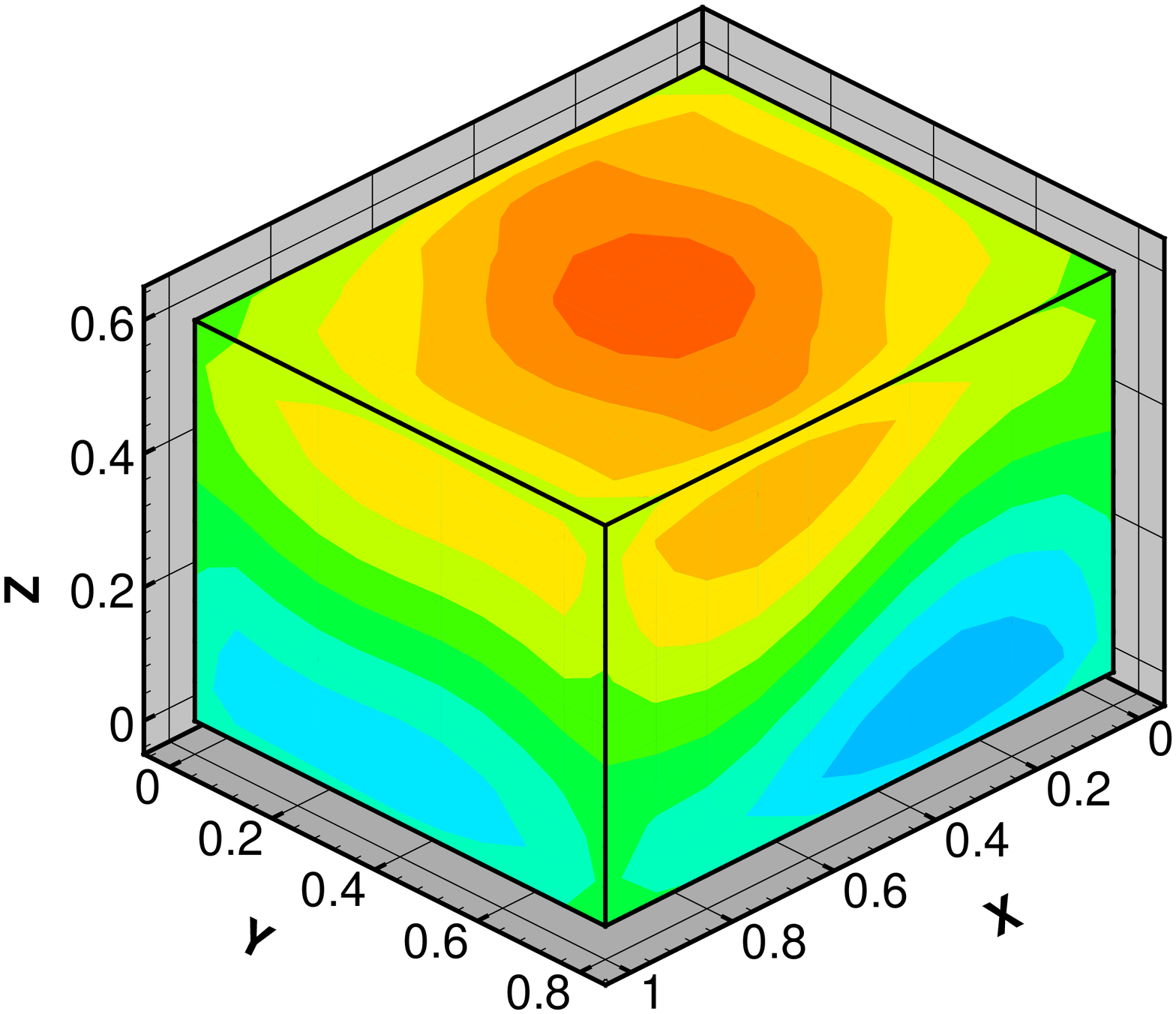}}\\
{\includegraphics[width=7cm]{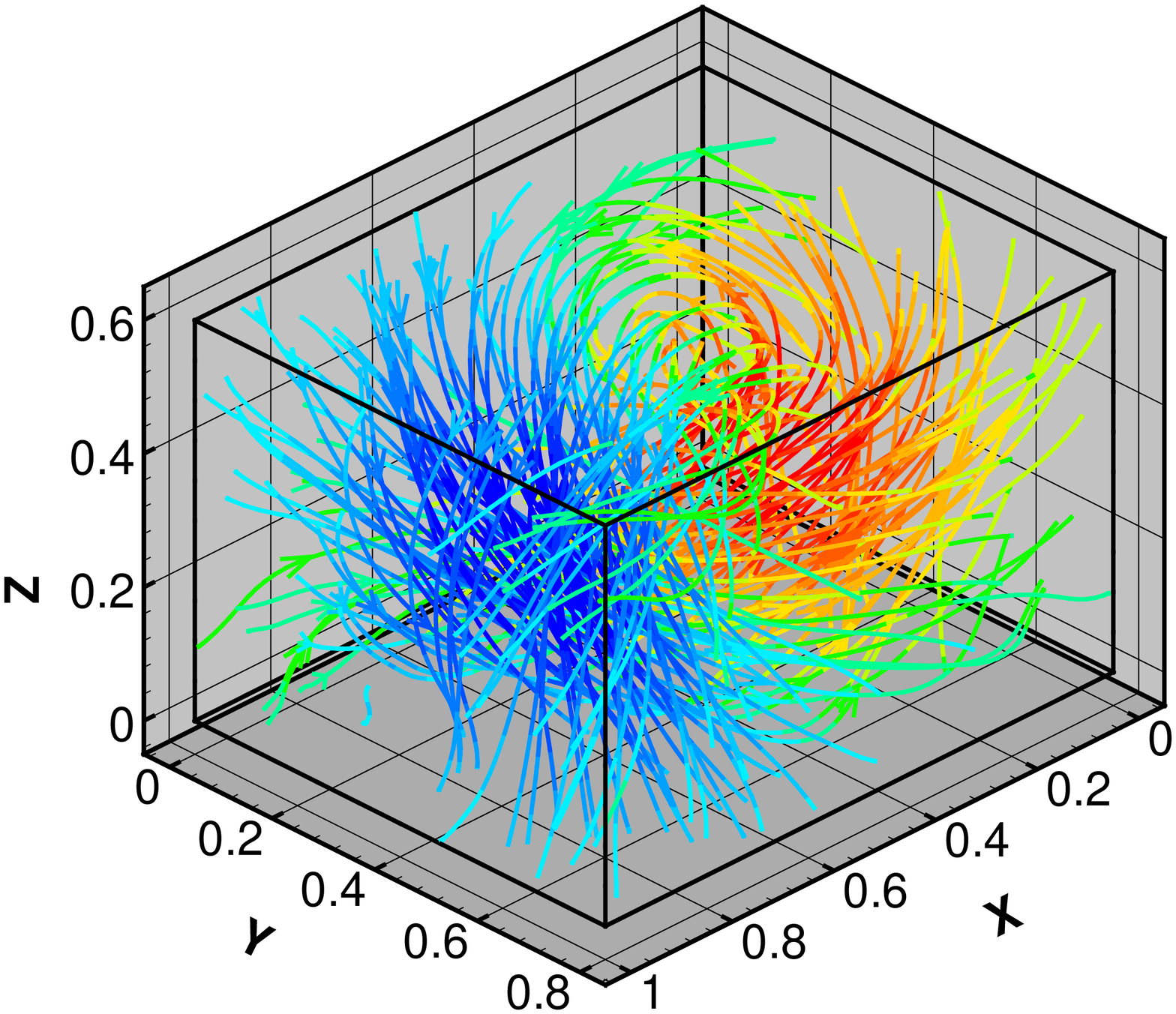}}
{\includegraphics[width=7cm]{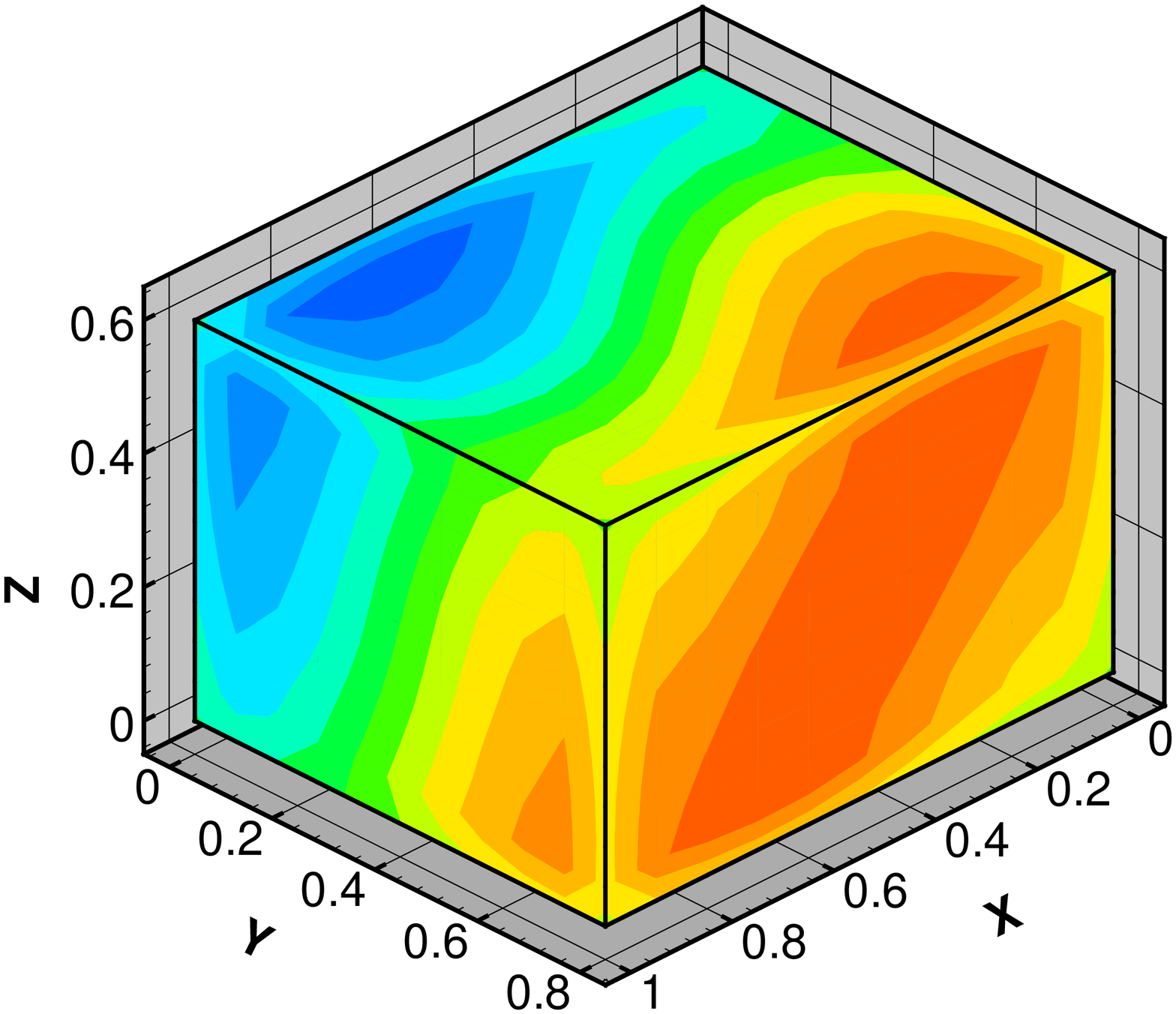}}\\
{\includegraphics[width=7cm]{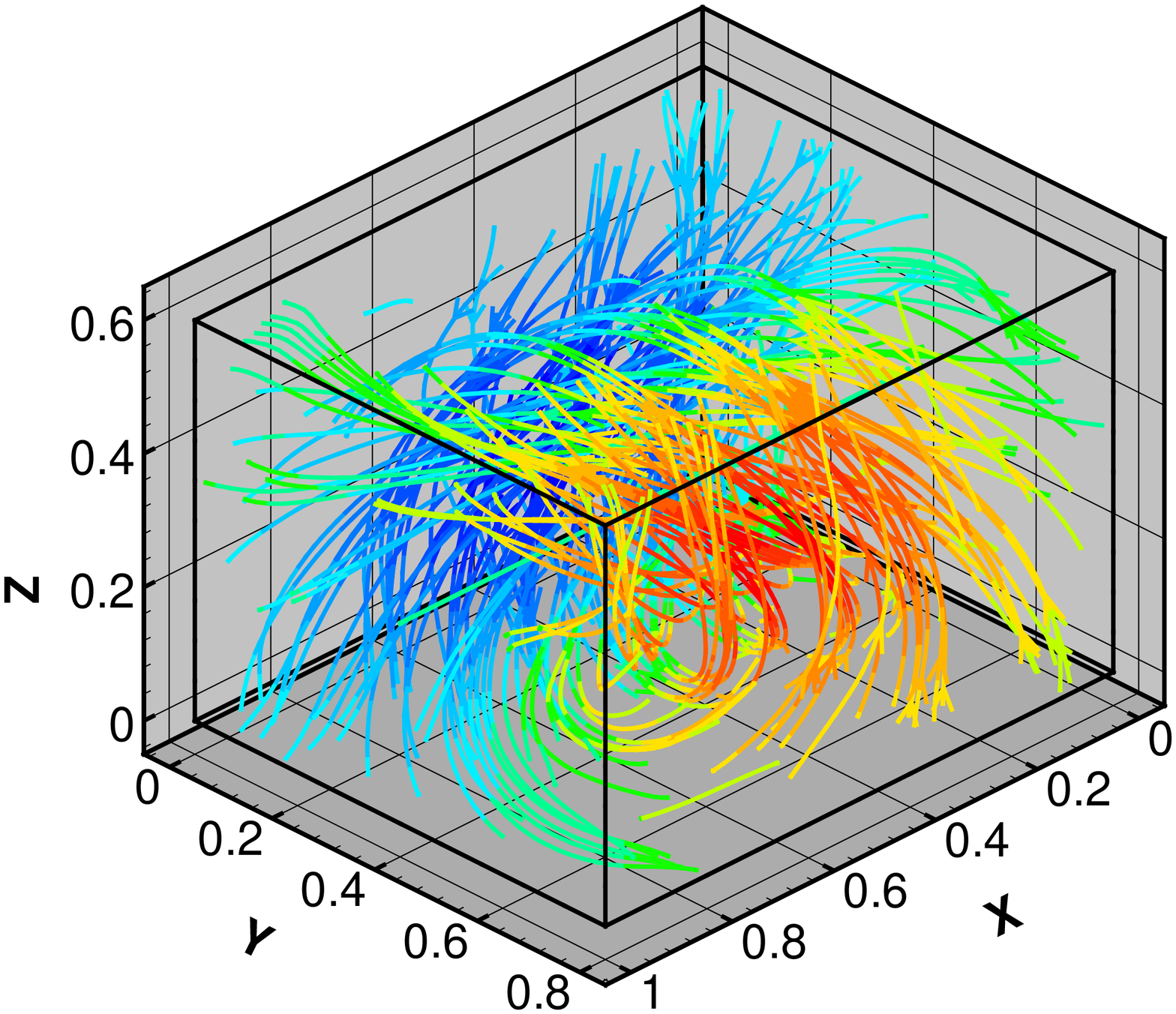}}
{\includegraphics[width=7cm]{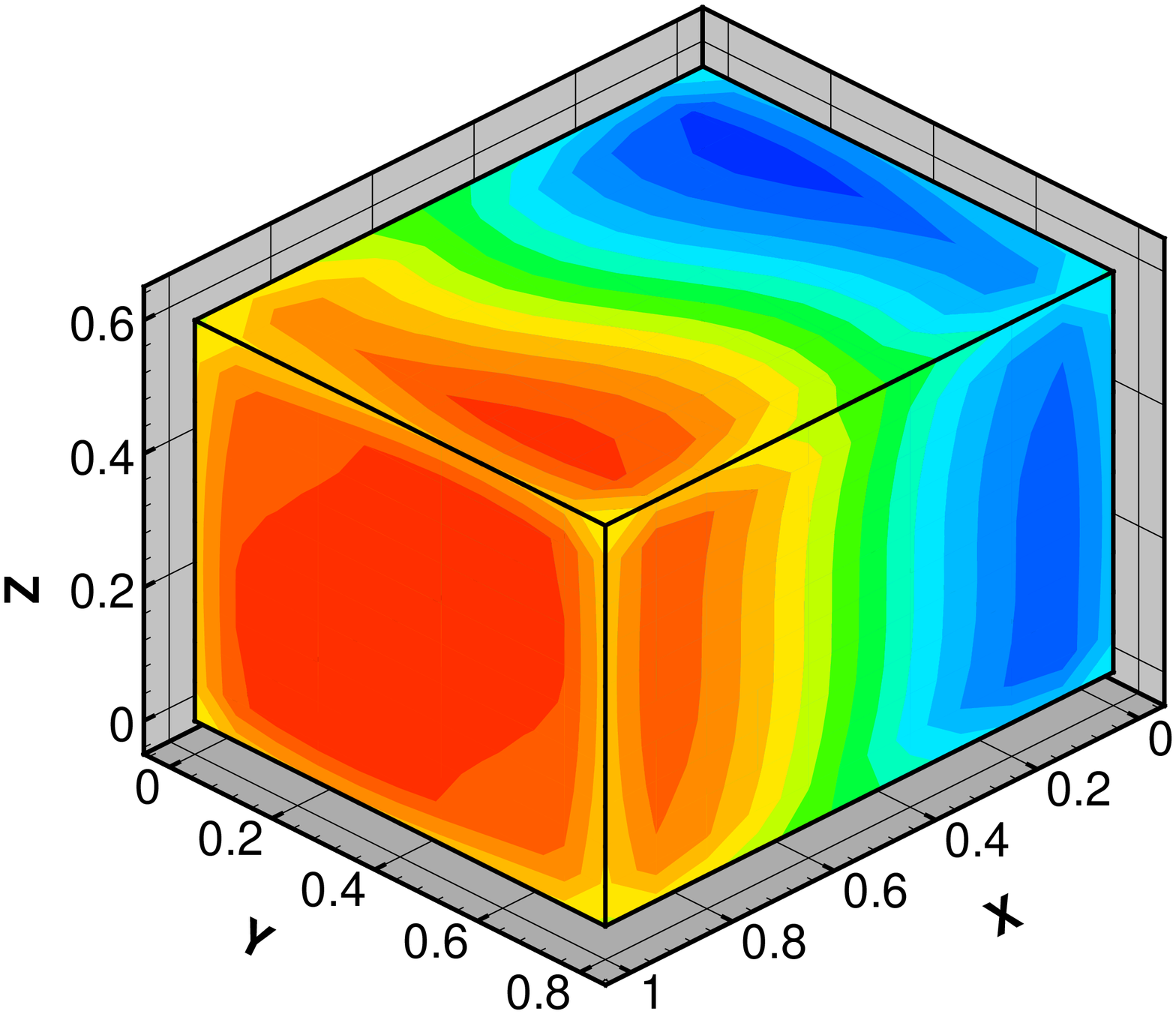}}
\caption{Magnetic fields within the box (left) and electric potentials at the
box boundary (right) belonging to the three  lowest
eigenvalues of the 
$\alpha^2$ dynamo in a box with sidelengths ratio 1:0.8:0.6. 
The corresponding eigenvalues of $C$ are 4.728 (top), 4.898 (middle), and
4.934 (bottom). Evidently, the eigenmode with the dipole axis
perpendicular to the largest box face (top) is most easily excitable.}
\label{fig9}
\end{figure}

For the particular case of a rectangular 
box with the sidelengths ratio 1.0:0.8:0.6 and 
a grid number $N=11$ we
have visualized the magnetic fields and electric potentials belonging to the 
three lowest eigenvalues 
(Fig. \ref{fig9}). The structure of the magnetic field with its typical 
mixture of poloidal and toroidal components
is clearly visible. It becomes evident  
that the field with the dipole axis perpendicular to the largest box face
has the lowest eigenvalue.

It should be noted that we have also checked the divergence-free condition of the 
fields and the curl-free condition in the exteriour. Both conditions are
fulfilled by the integral equation method with 
a reasonable accuracy.

\section{Conclusions}

We have used the integral equation method 
to solve numerically the steady
kinematic $\alpha^2$ dynamo problem in finite 
domains. 

For spherical domains, our approach is similar to the integral 
equation method for the 
solution of the radial Schr\"odinger equation.
In this case, with only some tens of grid points the method
provides reasonable results for all three considered example profiles
$\alpha(r)$. 
The error decreases with the inverse squared number of grid points. 
With the use of a 
convergence accelerating strategy, the accuracy and the convergence 
rate can be 
significantly improved.
Interestingly, even oscillating solutions of the dynamo problem
which cannot be reproduced by our steady method are at 
least mirrored by complex  eigenvalues for the dynamo numbers
whose real part is close to the correct critical value.

The particular suitability of the method to handle dynamos
in arbitrarily shaped domains was demonstrated by the treatment of an
$\alpha^2$ dynamo in rectangular boxes. 

In summary, the integral equation method seems to be 
an attractive tool for
the treatment of hydromagnetic dynamo problems.
The robustness and accuracy of the method encourages 
to generalize it to the unsteady case and to more complicated 
dynamo models.

\section{Acknowledgements}
This work was supported by Deutsche Forschungsgemeinschaft in frame
of the Collaborative Research Center SFB 609.

\section{Appendix: The handling of the singularities}
In this appendix, we present the techniques to handle 
the singularities in the integrals appearing in Eqs. (\ref{eq3}) and (\ref{eq4}) 
in general and for the matchbox in particular.
For the integral 
\[
\int_D
\frac{ {\mathbf{F}}  ({\mathbf{r'}}) \times 
({\mathbf{r}}-{\mathbf{r'}})}{|{\mathbf{r}}-{\mathbf{r'}}|^3} \; 
dV',
\]
with ${\mathbf{r}}$ belonging to $D$, we split the domain $D$ into 
two parts, one being a small subdomain $D_\epsilon$ which is usually 
defined as a ball of a radius $\epsilon$  centered 
at ${\bf{r}}$, the remaining being the region $D-D_\epsilon$. 
Hence the integral can be decomposed according to
\begin{eqnarray}{\label{adeqq}}
\int_D
\frac{ {\mathbf{F}}  ({\mathbf{r'}}) \times 
({\mathbf{r}}-{\mathbf{r'}})}{|{\mathbf{r}}-{\mathbf{r'}}|^3} \; 
dV'&=&\int_{D-D_{\epsilon}}\frac{ {\mathbf{F}}  ({\mathbf{r'}}) \times 
({\mathbf{r}}-{\mathbf{r'}})}{|{\mathbf{r}}-{\mathbf{r'}}|^3} \; 
dV'\nonumber 
\\ &&+\int_{D_\epsilon}\frac{ {\mathbf{F}}  ({\mathbf{r'}}) \times 
({\mathbf{r}}-{\mathbf{r'}})}{|{\mathbf{r}}-{\mathbf{r'}}|^3} \; 
dV'.
\end{eqnarray}
The first integral of the right hand side of this equation is a 
normal integral without any singularity. For the second integral, 
the introduction of the spherical coordinates system 
$(\rho, \theta, \phi)$ leads to
\begin{eqnarray}{\label{ad2a0}}
\int_{D_\epsilon}\frac{ {\mathbf{F}}  ({\mathbf{r'}}) \times 
({\mathbf{r}}-{\mathbf{r'}})}{|{\mathbf{r}}-{\mathbf{r'}}|^3} \; 
dV'=&\int_0^\epsilon\int_0^\pi\int_0^{2\pi}{\bf{F}}
(\rho,\theta,\phi)\times (\sin \theta \; \cos \phi, \sin \theta 
\; \sin \phi, \cos \theta)^T \;\nonumber\\
&\sin\theta \; d\phi\; d\theta\; d\rho
\end{eqnarray}
Assuming that the function ${\bf{F}}(\rho, \theta, \phi)$ is 
finite  
we see that the right hand side of Eq. (\ref{ad2a0})
vanishes in the limit $\epsilon\rightarrow 0$.

Therefore,  the considered singularity is a weak 
singularity. In order to avoid such a singularity in the 
numerical computation, we can just discretize the region 
$D-D_\epsilon$ instead of $D$. The error caused by this procedure 
can be made as small as desired by taking a 
small enough value of $\epsilon$. The same procedure can be applied
to the first integral on the right hand side of Eq. (\ref{eq4}) in the case 
$\bf{s}=\bf{r'}$. 

For the integral 
\begin{eqnarray}{\label{ad2a}}
\int\limits_S \varphi({\mathbf{s'}})
{\mathbf{n}}({\mathbf{s'}}) 
\cdot \frac{{\mathbf{s}}-{\mathbf{s'}}}
{{|{\mathbf{s}}-{\mathbf{s'}}|}^3} 
\;  dS', \nonumber
\end{eqnarray}
appearing in Eq. (\ref{eq4}),
with ${\mathbf{s}}$ on the surface $S$,
note that the unit vector ${\mathbf{n}}({\mathbf{s}}')$ 
tends to be perpendicular to the vector ${\mathbf{s}}-
{\mathbf{s}}'$ in the case that ${\mathbf{s}}\rightarrow 
{\mathbf{s}}'$, that is, ${\mathbf{n}}({\mathbf{s}}')
\cdot ({\mathbf{s}}-{\mathbf{s}}')$ tends to be zero. 
If defining $S_\epsilon$ as a small surface of a size 
$\epsilon$ including the point ${\mathbf{s}}$, we obtain \cite{BREB} (p. 69):
\begin{eqnarray}{\label{ad2b}}
\lim_{\epsilon\to 0} \int\limits_{S_\epsilon} \varphi({\mathbf{s'}})
{\mathbf{n}}({\mathbf{s'}}) 
\cdot \frac{{\mathbf{s}}-{\mathbf{s'}}}{{|{\mathbf{s}}-{\mathbf{s'}}|}^3} 
\;  dS'\rightarrow 0.
\end{eqnarray}
A similar 
strategy as mentioned above can be employed to avoid such a singularity by 
discretizing the surface $S-S_\epsilon$ instead of $S$.

Now we consider the integral
\begin{eqnarray}{\label{ad2c}}
 \int_S \varphi({\mathbf{s'}}) 
{\mathbf{n}} ({\mathbf{s'}}) \times
\frac{{\mathbf{r}}-{\mathbf{s'}}}{|{\mathbf{r}}-{\mathbf{s'}}|^3} 
\; dS',
\end{eqnarray}
with the point ${\mathbf{r}}$ sitting on the surface $S$. Denote $S_\epsilon$ as a 
small piece of the surface $S$, which satisfies 
$|{\mathbf{s}}'-{\mathbf{r}}|<\epsilon$. 
Then the integral (\ref{ad2c}) can be written as
\begin{eqnarray}{\label{ad2d}}
\int_S \varphi({\mathbf{s'}}) 
{\mathbf{n}} ({\mathbf{s'}}) \times
\frac{{\mathbf{r}}-{\mathbf{s'}}}{|{\mathbf{r}}-{\mathbf{s'}}|^3} 
\; dS'=&\int_{S-S_{\epsilon}} \varphi({\mathbf{s'}}) 
{\mathbf{n}} ({\mathbf{s'}}) \times
\frac{{\mathbf{r}}-{\mathbf{s'}}}{|{\mathbf{r}}-{\mathbf{s'}}|^3} 
\; dS'\nonumber\\
&+\int_{S_{\epsilon}} \varphi({\mathbf{s'}}) 
{\mathbf{n}} ({\mathbf{s'}}) \times
\frac{{\mathbf{r}}-{\mathbf{s'}}}{|{\mathbf{r}}-{\mathbf{s'}}|^3} 
\; dS'.
\end{eqnarray}
The first integral of the right hand side of this equation is a normal integral with 
no singularity. As for the second, defining another small disk of a small enough 
radius $\eta$ centered at $\mathbf{r}$, we have
\begin{eqnarray}
\lim_{\epsilon\to 0}\int_{S_{\epsilon}} \varphi({\mathbf{s'}}) 
{\mathbf{n}} ({\mathbf{s'}}) \times
\frac{{\mathbf{r}}-{\mathbf{s'}}}{|{\mathbf{r}}-{\mathbf{s'}}|^3} 
\; dS'&=&\lim_{\epsilon\to 0}\;
\lim_{\epsilon\to 0} \int_{S_{\epsilon}-S_{\eta}}\varphi({\mathbf{s'}}) 
{\mathbf{n}} ({\mathbf{s'}}) \times
\frac{{\mathbf{r}}-{\mathbf{s'}}}{|{\mathbf{r}}-{\mathbf{s'}}|^3} 
\; dS'\nonumber\\
&=&\varphi({\mathbf{r}}) 
{\mathbf{n}} ({\mathbf{r}})\times \lim_{\epsilon\to 0}\;
\lim_{\eta \to 0}\int_{S_{\epsilon}-S_{\eta}}
\frac{{\mathbf{r}}-{\mathbf{s'}}}{|{\mathbf{r}}-{\mathbf{s'}}|^3} 
\; dS',\nonumber
\end{eqnarray}
where $0<\eta<\epsilon$. Introducing the local polar coordinates, $dS'=\rho d\theta d\rho$, leads to
\begin{eqnarray}{\label{ad2f}}
\lim_{\eta\to 0}\int_{S_{\epsilon}-S_{\eta}}
\frac{{\mathbf{r}}-{\mathbf{s'}}}{|{\mathbf{r}}-{\mathbf{s'}}|^3} 
\; dS'&=&-\lim_{\eta\to 0} \int_\eta^\epsilon\frac{1}{\rho}d\rho 
\int_0^{2\pi}(\cos\theta, \sin\theta)^Td\theta\nonumber\\
&=&-\lim_{\eta\to 0} \ln \frac{\epsilon}{\eta}\int_0^{2\pi}
(\cos\theta, \sin\theta)^T d\theta.
\end{eqnarray}
Since the integrals $\int_0^{2\pi}\cos \theta \; d\theta$ and $\int_0^{2\pi}\sin\theta 
 \; d\theta$ always vanish, we have
\begin{eqnarray}{\label{ad2g}}
\lim_{\eta \to 0}\int_{S_{\epsilon}-S_{\eta}}
\frac{{\mathbf{r}}-{\mathbf{s'}}}{|{\mathbf{r}}-{\mathbf{s'}}|^3} 
\; dS'=0.
\end{eqnarray}
Therefore,
\begin{eqnarray}{\label{ad2h}}
\lim_{\epsilon \to 0}\int_{S_{\epsilon}} \varphi({\mathbf{s'}}) 
{\mathbf{n}} ({\mathbf{s'}}) \times
\frac{{\mathbf{r}}-{\mathbf{s'}}}{|{\mathbf{r}}-{\mathbf{s'}}|^3} 
\; dS'=0.
\end{eqnarray}
This indicates that we can also discretize the surface $S-S_\epsilon$ 
instead of $S$ in order to avoid such a singularity in the numerical computation. 

For more details of the handling of the various singularities, 
one may refer to \cite{PARI} (pp. 7-16).


\begin{thebibliography}{99}
\bibitem{ATKI} K. Atkinson, {\it The Numerical Solution of 
Integral Equations of the Second Kind} (Cambridge University Press, Cambridge, 1997).
\bibitem{BARN}A. C. L. Barnard, I. M. Duck, M. S. Lynn, 
and W. P. Timlake, The application of electromagnetic theory to 
electrocardiography. II. Numerical solution of the integral equations, 
{\it Biophys. J.} {\bf 7}, 463-491 (1967).  
\bibitem{BREB}C. A. Brebbia, J. C. F. Telles and L. C. Wrobel, 
{\it Boundary Element Techniques} (Springer, Berlin, 1984).
\bibitem{BUEN} E. Buendia, R. Guardiola, and M. Montoya, Integral Equations:
A Tool to Solve the Schr\"odinger Equation,
{\it J. Comput. Phys.} {\bf 68}, 187-201 (1987).
\bibitem{BRAN} A. Brandenburg, \AA. Nordlund, R. F. Stein,  and 
Torkelsson, Dynamo-generated turbulence and large scale
magnetic fields in a Keplerian shear flow,  {\it ApJ} {\bf{446}}, 741--754 (1995).
\bibitem{DORA} W. Dobler and K.-H. R\"adler,  An integral equation 
approach to kinematic dynamo models, {\it Geophys. 
Astrophys. Fluid Dyn.} {\bf{89}}, 45-74 (1998).
\bibitem{CHRI} T. H. Christopher and M. A.  Baker, {\it The Numerical 
Treatment of Integral Equations} (Clarendon Press, Oxford, 1977).
\bibitem{DELV1} L. M. Delves and J. Walsh, {\it Numerical Solution of 
Integral Equations} (Clarendon Press, Oxford, 1974).
\bibitem{DELV2} L. M. Delves and J. L. Mohamed, {\it 
Computional methods 
for integral equations} (Cambridge University Press, Cambridge, 1985).
\bibitem{FREI} Ya. Freiberg, Optimization of the shape of the toroidal
model of an MHD dynamo, {\it Magnetohydrodynamics}  {\bf{11}}, No 3, 269-272 (1975).
\bibitem{GAI2} A. Gailitis, Self-excitation of a magnetic field
by a pair of annular vortices, {\it Magnetohydrodynamics} {\bf 6}, No 1, 14-17 (1970). 
\bibitem{GAI3} A. Gailitis and Ya. Freiberg, Self-excitation of a 
magnetic field by a pair of annular eddies, {\it Magnetohydrodynamics} {\bf 10}, No 1,
26-30 (1974).
\bibitem{GAI4} A. Gailitis, O. Lielausis, S. Dement'ev, E. Platacis, A. Cifersons, 
G. Gerbeth, Th. Gundrum, F. Stefani, M. Christen, H. H\"anel, and G. Will, 
Detection of a flow induced magnetic field eigenmode in the Riga dynamo facility,
{\it Phys. Rev. Lett.} {\bf 84}, 4365-4368 (2000).
\bibitem{GAI5} A. Gailitis, O. Lielausis, E. Platacis, S. Dement'ev, A. Cifersons, 
G. Gerbeth, Th. Gundrum, F. Stefani, M. Christen, and G. Will,
Magnetic field saturation in the Riga dynamo experiment,
{\it Phys. Rev. Lett.} {\bf 86}, 3024-3027 (2001).
\bibitem{GAI6} A. Gailitis, O. Lielausis, E. Platacis, G. Gerbeth, and 
F. Stefani., Laboratory Experiments on Hydromagnetic Dynamos,
{\it Rev. Mod. Phys.} {\bf 74}, 973-990 (2002).
\bibitem{GON1} R. A. Gonzales, J. Eisert, I. Koltracht, M. Neumann, and
G. Rawitscher, Integral equation method for the continuous
spectrum radial Schr\"odinger
equation, {\it J. Comp. Phys.} {\bf 134}, 134-149 (1997).
\bibitem{GON2} R. A. Gonzales, S.-Y. Kang, I. Koltracht, and G. Rawitscher, 
Integral equation method for coupled Schr\"odinger equations,
{\it J. Comp. Phys.} {\bf 153}, 160 - 202 (1999).
\bibitem{GRRO} L. Greengard and V. Rokhlin,
On the Numerical Solution of Two-Point Boundary Value Problems,
{\it Comm. Pure Appl. Math} {\bf 44}, 
419--452 (1991).
\bibitem{HACK} W. Hackbusch, {\it Integral Equations: Theory and 
Numerical Treatment} (Springer, Berlin, 1995).
\bibitem{HAEM} M. H\"am\"al\"ainen, R. Hari, R. J. Ilmoniemi, J. Knuutila, 
O. V. Lounasmaa, Magnetoencephalography - theory, instrumentation, 
and application to noninvasive studies of the working human brain,
{\it Rev. Mod. Phys.} {\bf 65},  413--497 (1993).
\bibitem{KRRA} F. Krause and K.-H. R\"adler, {\it Mean-field 
Magnetohydrodynamics
and Dynamo Theory} (Akademie-Verlag, Berlin, 1980).
\bibitem{KRST} F. Krause and M. Steenbeck, 
Untersuchung der Dynamowirkung einer nichtspiegelsymmetrischen
Turbulenz an einfachen Modellen, {\it  Z. Naturforsch.} {\bf 22a}, 671-675 (1967).
\bibitem{MES2} A. J. Meir and P. G. Schmidt, 
Analysis and numerical approximation of a stationary 
MHD flow problem with non-ideal boundary,
{\it SIAM J. Numer. Anal.} {\bf 36}, 
1304 (1999).
\bibitem{MOFF} H. K. Moffatt, {\it Magnetic field generation in electrically 
conducting fluids} (Cambridge University Press, Cambridge, 1978).
\bibitem{MUST} U. M\"uller and  R. Stieglitz, 
Can the Earth's magnetic field be simulated in the laboratory?,
{\it Naturwissenschaften} {\bf 87}, 381-390 (2000).
\bibitem{PARI}F. Par\'{i}s and J. Ca\~{n}as, {\it Boundary 
Element Method (Fundamentals and Applications)} (Oxford University Press, 1997).
\bibitem{PTVF} W. H. Press, S. A.  Teukolsky, W. T.  Vetterling,
B. F. Flannery, {\it Numerical Recipes} (Cambridge University 
Press, Cambridge, 1992).
\bibitem{RAE1} K.-H. R\"adler, E.  Apstein,  M. Rheinhardt, and 
M. Sch\"uler, The Karlsruhe dynamo experiment - a mean-field
approach, {\it Studia geoph. et geod.} {\bf 42}, 224-231 (1998).
\bibitem{RAE2}
K.-H. R\"adler, M. Rheinhardt, E. Apstein, and H. Fuchs, 
On the mean-field theory of the Karlsruhe dynamo experiment, 
{\it Nonlinear  Proc. Geoph.} {\bf 9}, 171-187 (2002)
\bibitem{ROB1} P. H. Roberts, {\it An Introduction to 
Magnetohydrodynamics} (Elsevier, New York, 1967).
\bibitem{RUZH} G. R\"udiger and Y.  Zhang,
MHD instability in differentially-rotating cylindric flows.
{\it Astron. Astrophys.} {\bf 378}, 302-308 (2001).
\bibitem{STE1} F. Stefani, G. Gerbeth, and A. Gailitis, 
Velocity profile optimization for the Riga dynamo experiment,
in {\it Transfer Phenomena in Magnetohydrodynamics and 
Electroconducting Flows}, edited by A. Alemany, Ph. Marty, and J.-P. Thibault,
31-44 (Kluwer, Dordrecht, 1999).
\bibitem{STE2} F. Stefani, G.  Gerbeth, and K.-H. R\"adler, 
Steady dynamos in finite domains: an integral equation approach,
{\it Astron. Nachr.} {\bf 321}, 65-73 (2000).
\bibitem{STE3} F. Stefani and G. Gerbeth, G.,
Oscillatory mean-field dynamos with a spherically symmetric,    
isotropic helical turbulence parameter $\alpha$,
{\it Phys. Rev. E}  {\bf 67}, 027302 (2003)
\bibitem{STMU} R. Stieglitz and U. M\"uller, U.,
Experimental demonstration of a homogeneous two-scale 
dynamo, {\it Phys. Fluids} {\bf 13}, 561-564 (2001).
\end{thebibliography}
\end{document}